\documentclass[journal]{IEEEtran}
\usepackage{amsmath,amsfonts,amssymb}
\usepackage{algorithmic}
\usepackage{algorithm}
\usepackage{array}
\usepackage[caption=false,font=normalsize,labelfont=sf,textfont=sf]{subfig}
\usepackage{textcomp}
\usepackage{stfloats}
\usepackage{url}
\usepackage{verbatim}
\usepackage{graphicx}
\usepackage{cite}

\usepackage{setspace}
\usepackage{textcomp}
\usepackage{booktabs,multirow}
\usepackage{tabularx}
\usepackage[table,dvipsnames,svgnames,x11names]{xcolor}
\usepackage{colortbl}
\usepackage{hhline}
\usepackage{arydshln}
\usepackage{relsize}
\usepackage{float}
\usepackage{threeparttable}
\usepackage{pifont}
\usepackage{marvosym}
\usepackage{ragged2e}
\usepackage{multirow}
\usepackage{vcell}
\usepackage{courier}
\usepackage{booktabs}
\usepackage{soul}
\usepackage{comment}
\usepackage[hidelinks,colorlinks=true,linkcolor=Emerald,citecolor=cyan,urlcolor=black]{hyperref}
\usepackage{soul}

\usepackage{pgfplots}
\usepgfplotslibrary{groupplots}
\pgfplotsset{compat=1.18}
\usepgfplotslibrary{statistics}

\usepackage{tikz,xcolor}

\definecolor{lime}{HTML}{A6CE39}
\DeclareRobustCommand{\orcidicon}{%
	\begin{tikzpicture}
	\draw[lime, fill=lime] (0,0) 
	circle [radius=0.14] 
	node[white] {{\fontfamily{qag}\selectfont \tiny ID}};
	\draw[white, fill=white] (-0.0625,0.095) 
	circle [radius=0.007];
	\end{tikzpicture}
	\hspace{-2mm}
}

\foreach \x in {A, ..., Z}{%
	\expandafter\xdef\csname orcid\x\endcsname{\noexpand\href{https://orcid.org/\csname orcidauthor\x\endcsname}{\noexpand\orcidicon}}
}

\newcommand{\ignore}[1]{}

\definecolor{mysecpurple}{RGB}{115,30,128}

\begin{document}

%From Coin Flips to Controlled Probabilities: True-Random Bit-Stream Mapping for Stochastic Computing
%\huge From Entropy Source to Stochastic Bit-Stream: Technology-Agnostic Converters for True-Random Stochastic Computing
\title{From Entropy Source to Stochastic Bit-Stream: Low-Cost Converters for True-Random\\ Stochastic Computing}
%TRNG-to-SNG: Trading Determinism for Entropy\\ in Stochastic Number Generation
%\huge Trading Determinism for Entropy:\\ Low-Cost TRNG-to-SNG for Stochastic Computing
%\vspace{-0.5em}

%\author{}

\author{Mehran~Moghadam\orcidA{}, \IEEEmembership{Graduate Member,~IEEE}, Abu Masum\orcidB{}, \IEEEmembership{Graduate Member,~IEEE}, M.~Hassan~Najafi\orcidD{}, \IEEEmembership{Senior Member,~IEEE}, Sercan~Aygun\orcidC{}, \IEEEmembership{Senior Member,~IEEE}, and Di Wu\orcidE{}, \IEEEmembership{Member,~IEEE}
\thanks{This work is supported by NSF grants 2339701, 2609436, 2616837, NASA grant 80NSSC25C0335, Vernon $\&$ Ruby Langlinais Non-Endowed Research Fund, Lockheed Martin Corporation Endowed Professorship Fund, and gifts from NVIDIA and Google. Mehran Moghadam and M. Hassan Najafi are with the ECSE %Electrical, Computer, and Systems Engineering 
Department, Case Western Reserve University, Cleveland, OH, USA. E-mail:\{moghadam, najafi\}@case.edu. Abu Masum and Sercan Aygun are with the School of Computing and Informatics, University of Louisiana at Lafayette, Lafayette, LA, USA. E-mail:\{c00591145, sercan.aygun\}@louisiana.edu. Di Wu is with the Department of ECE at the University of Central Florida. E-mail: di.wu@ucf.edu}% <-this % stops a space
%\thanks{\scriptsize A preliminary version of this work appeared as~\cite{ID_VSA_ISLPED25}.}
%\vspace{-5pt}
%\vspace{-2.em}
}

%\markboth{IEEE Journal on Emerging and Selected Topics in Circuits and Systems Vol.~XX, No.~X, XXXX~2026}%


\maketitle

\begin{abstract}
Stochastic computing (SC) encodes values as the probability of \textit{ones} in random bit-streams, 
which allows arithmetic circuits with very low area and power and a natural tolerance to faults. The accuracy and security of an SC system depend on the quality of those bit-streams. 
Conventional generators build them from pseudo-random or quasi-random sequences, which are predictable, periodic, and correlated across streams drawn from the same generator.
%enabling extremely simple arithmetic circuits with low area, low power, and inherent resilience to faults. %minimal hardware arithmetic and substantial reductions in area and power, along with inherent fault tolerance.
%However, the effectiveness of SC critically depends on the statistical quality of the underlying bit-streams.
%The effectiveness of SC, however, is tightly coupled to the quality of bit-streams. 
%Conventional bit-stream generators, typically built on pseudo-random or quasi-random sequences, suffer from deterministic patterns, bias, and limited unpredictability, which can degrade accuracy and make them less suitable for applications requiring stronger robustness and security.
%Conventional bit-stream generators, typically based on pseudo-random number generators (PRNGs) such as linear-feedback shift registers or quasi-random number generators (QRNGs) such as Sobol and Halton sequences, suffer from bias and predictable patterns that %. These weaknesses 
%not only degrade accuracy but also limit their suitability in applications that require strong unpredictability.
This work proposes three methods for converting the unbiased outputs of a true random number generator (TRNG) into stochastic bit-streams (SBS) with arbitrary target probabilities: a \textit{block extraction} and comparison circuit, a \textit{counter-logic} circuit, and a cascaded \textit{majority-logic} pipeline. All three accept bits from any true-random source and are built from standard CMOS logic.
%Unlike prior deterministic generators, the proposed converters directly leverage \textit{true-random} inputs while remaining fully compatible with standard CMOS implementation. %Together, these approaches eliminate reliance on deterministic sequences 
%while offering accuracy, robustness, and low implementation costs.
%Comprehensive evaluations show that the proposed converters generate high-quality SBS with accuracy close to software-based random generation for both independent and correlated bit-stream generation.
They reach the accuracy of a software random generator for both independent and correlated stream generation, and they support the common SC arithmetic operations with error that falls with stream length at the same rate.
%Evaluation results show that the proposed designs achieve high-quality bit-stream generation with improved efficiency and enhanced resilience compared to state-of-the-art PRNG/QRNG-based designs.
%Hardware synthesis results demonstrate favorable cost-efficiency, with compact area, low energy, and attractive delay–area trade-offs relative to representative pseudo- and quasi-random baselines. 
Hardware synthesis results show that the converters offer favorable area, energy, and area-delay trade-offs relative to pseudo- and quasi-random baselines.
Under injected soft errors, block extraction degrades no faster than the best deterministic baseline. 
By bridging raw entropy sources and application-ready bit-streams, this work makes true-random SC practical on standard CMOS.
\end{abstract}


\vspace{-0.15em}
\section{Introduction}
%\vspace{-0.15em}
\IEEEPARstart{S}{tochastic} computing (SC) has attracted renewed attention as an unconventional %yet powerful 
computing paradigm with distinct advantages over traditional binary arithmetic~\cite{arminSurvey, WeikangSurvey, LiSurvey}. Instead of representing a number by weighted binary digits, SC encodes it as the probability of \textit{ones} in an $N$-bit random bit-stream. This probability-based representation lets very simple hardware perform complex operations, often with just a few logic gates.
%enables implementing complex operations with extremely simple hardware, often requiring just a handful of logic gates. 
Multiplication, for example, is %complex and 
costly in binary radix but needs only a single \texttt{AND} gate in SC~\cite{Alaghi_SCC}. 
SC circuits therefore occupy far less area and consume far less power than their binary counterparts. 
%The result is a significant reduction in circuit area and power consumption, along with inherent fault and noise tolerance, as errors in individual bits have a minimal impact on overall accuracy. 
They also tolerate high rates of faults and noise, since a flipped bit shifts the encoded value only slightly.
These properties make SC attractive for error-prone and resource-constrained applications, including deep learning accelerators~\cite{ubrain_ISCAS22}, digital signal processing~\cite{10041889}, image processing~\cite{PengRiedelImageProcessingCaseStudies}, and bio-science~\cite{electronics10232981}, where compactness, parallelism, and robustness matter more than exact precision. SC has also been investigated in emerging computing contexts such as neuromorphic systems~\cite{9639029} and in-memory architectures~\cite{RiahiAlam_TCAS2022,MITRA_ISLPED2026,FALCON_arXiv2026}, underscoring its potential as a low-cost foundation for next-generation hardware.

A central challenge in SC is generating 
high-quality bit-streams. The accuracy, reliability, and efficiency of SC operations depend on how  \textit{random} these streams are and how they \textit{correlate} with one another. Conventional stochastic bit-stream (SBS) generators rely on \textit{pseudo-random} number generators (PRNGs), such as linear-feedback shift registers (LFSRs), or \textit{quasi-random} number generators (QRNGs), such as Sobol~\cite{liu2017energy}, Halton~\cite{Alaghi_Fast}, and van der Corput (VDC)~\cite{p2lsg_ASP_DAC24,TranSC_JETCAS2026} sequences. 
These generators are attractive because their behavior is well understood and their hardware is mature. They are also deterministic. Their outputs follow \textit{fixed patterns}, \textit{repeat} after a finite period, and \textit{correlate} with one another when streams share a generator.
%While attractive for their well-understood behavior and mature hardware support, these algorithmic approaches inevitably suffer from \textit{predictable structure}, \textit{correlations}, and \textit{periodicity} that degrade the statistical fidelity of the generated bit-streams. 
%This becomes especially problematic in security- and privacy-sensitive applications, where the deterministic nature of PRNG/QRNG outputs makes them vulnerable to reverse engineering and adversarial exploitation. 
In security- and privacy-sensitive applications this determinism becomes a liability: an adversary who knows the generator can reproduce or reverse engineer the streams. 
Even recent %state-of-the-art (SOTA) 
designs that integrate SC into in-memory architectures and reach high throughput with little data movement, inherit this limitation because they still depend on algorithmically generated sequences~\cite{RiahiAlam_TCAS2022, RiahiAlam_DT_2021}. %As a result, 
What is missing is a conversion mechanism that takes bits from a \textit{true random} source and produces \textit{unbiased} bit-streams with a chosen probability, so that SC systems keep both their accuracy and their robustness.
%The community continues to face a growing need for conversion mechanisms that leverage \textit{true random} sources to produce \textit{high-quality}, \textit{unbiased}, and \textit{unpredictable} bit-streams, thereby preserving both computational integrity and the resilience of SC systems.

Unlike deterministic sources, true random number generators (TRNGs) produce sequences with high entropy and minimal correlation. This gives SC four advantages: I)~\textit{Accuracy}: 
SC computes on the statistics of bit-streams, so any bias or correlation in the random source appears directly as error in the result. Unbiased, uncorrelated bits remove this error source.
%High-quality randomness eliminates bias or unwanted correlation in bit-streams, reducing errors in probabilistic computations. Since SC relies on the statistical properties of bit-streams, bias or correlation in the random numbers can lead to significant errors.  
II)~\textit{Robustness}: a bit flip in a TRNG stream corrupts one bit. A bit flip in the state of an LFSR or Sobol generator changes every bit that follows.
%Resilience to noise and variability, in contrast to conventional PRNGs and QRNGs, which are built from components vulnerable to noise or bit-flip. 
III)~\textit{Efficiency}: 
conventional SC generators are expensive, often costing more than the arithmetic circuits they feed, which cancels much of the hardware saving that SC promises. A TRNG-based front end can cut this overhead in both logic and energy.
%Reduced logic overhead and energy consumption compared to PRNG/QRNG-based generators. Conventional SC bit-stream generators are notorious for their high cost, which offsets any hardware cost savings from the simplicity of the computation circuits. 
IV)~\textit{Security}: 
unpredictable streams resist reverse engineering and statistical attacks.
%Unpredictability mitigates vulnerabilities to reverse-engineering or statistical attacks. 
These advantages only materialize once a gap is closed. A TRNG produces \textit{coin flips}, with ones and zeros equally likely. SC needs bit-streams with \textit{arbitrary target probabilities}. 
%Despite these advantages, a key gap remains between \textit{unbiased coin flips} and \textit{controlled probabilities}: 
How to convert the first into the second efficiently is the question this work addresses.
%\textcolor{Navy}{\textit{How to efficiently convert unbiased random bits from TRNGs (50\% ones) into stochastic bit-streams (SBS) with arbitrary target probabilities?}
%} %This conversion is essential to encode real values in the $[0,1]$ interval into high-quality bit-streams to be used in SC systems.

%\red{[Maybe adding a discussion on those earlier works on SC bit-stream generation with controlling the write pulse]}

%There have been several prior efforts to generate 
Prior work has generated true random SBS by exploiting the stochastic behavior of emerging devices. Knag \textit{et al.}~\cite{SC_InMemory_Native2014} proposed a native SC architecture in which memristor arrays are programmed with group-write operations to produce SBS directly. Lammie \textit{et al.}~\cite{IMC_SC_DNN_2021} leveraged the probabilistic switching of Conductive Bridging RAM (CBRAM) devices to generate SBS for deep learning parameter optimization. Zhao \textit{et~al.}~\cite{Physic_RRAM_SC_SNG_2019} developed physics-based models of RRAM switching to generate SBS in memory. %demonstrated accurate in-memory bit-stream generation for SC systems. 
Gupta \textit{et~al.}~\cite{SCRIMP_DATE20} 
combined stochastic number generation and logic operations inside ReRAM crossbars. 
%proposed a ReRAM-based architecture that combines stochastic number generation and logic operations entirely inside memory crossbars. %In contrast to these prior works, the TRNG-to-SBS conversion methods proposed in this work are not dependent to % implemented fully in off-memory CMOS logic, providing TRNG-to-SBS conversion %true-random SBS generation 
%without dependence on 
%specific memory technologies. This independence makes our designs portable, broadly applicable, and easily integrable into standard CMOS digital systems. \blue{[We can discuss that all the proposed approaches are technology-agnostic (meaning that it is not tied to a specific technology like CMOS)]}
All of these generators are tied to a specific memory technology.
%In contrast to prior true-random SBS generators built around specific memory technologies, the TRNG-to-SBS conversion methods proposed in this work are technology-agnostic, broadly applicable, and readily integrable into standard CMOS digital systems.
The TRNG-to-SBS converters proposed here are not: they accept bits from any true random source and are built from standard CMOS logic, so they drop into existing digital systems without new device technology.

This work introduces 
%In this work, we introduce % address this gap by introducing 
three %novel 
%hardware 
methods for TRNG-to-SBS conversion: 1) \textbf{\textit{Block extraction \& comparison method}}: the TRNG outputs are split into $M$-bit blocks, each read as a random integer and compared against a stored target value. 2) \textbf{\textit{Counter-logic method}}: %A combinational structure employing a 
a down-counter, a multiplexer (\texttt{MUX}), and a few standard gates flip a controlled number of bits in the TRNG stream until the output reaches the target probability. 3) \textbf{\textit{Majority-logic pipeline method}}: a chain of majority gates, one stage per bit of the target value, biases the unbiased input toward the target probability with no comparator or counter.
%An architecture based on cascaded majority gates, offering a structurally simple yet effective means of converting unbiased inputs into probability-controlled streams.
%The proposed methods are designed entirely in CMOS logic without reliance on specific technologies, providing portable and efficient solutions. 
We evaluate the three methods for accuracy, speed, area, power, energy consumption, and soft-error tolerance, and compare them with state-of-the-art (SOTA) generators built on PRNGs and QRNGs. 
Our results show that the block-extraction and majority-logic converters occupy less area and use less energy per cycle than LFSR- and Sobol-based generators, and all three approach the accuracy of a software random generator as the bit-stream length grows.
%Our results demonstrate that the proposed designs enable true-random bit-stream generation at substantially lower cost and with stronger unpredictability compared to existing deterministic approaches, while maintaining high accuracy across a wide range of target probabilities. 
By closing the gap between raw entropy sources and application-ready SBS, this work makes true-random SC practical on standard CMOS.
%strengthens the practicality of secure and energy-efficient SC for next-generation computing systems.

The remainder of this paper is organized as follows. Section~\ref{Sec:ConvSNG} reviews the conventional design for SBS generation. Section~\ref{prop} presents the three proposed TRNG-to-SBS converters. Sections~\ref{perf} evaluates their accuracy, and Section~\ref{hard} compares their hardware cost with SOTA generators. Section~\ref{Sec:softerror} examines their tolerance to soft errors. Section~\ref{case_study} evaluates the bilinear image interpolation case study, and finally %\red{Section~\ref{Sec:CaseStudies} illustrates the performance for}  selected case studies. 
Section~\ref{conc} concludes the paper.

\section{Conventional SC Bit-stream Generation}
\label{Sec:ConvSNG}

%[Do we need a section on background on conventional methods for stochastic bit-stream generation? we may not]

\begin{figure}[t]
    \centering
    \includegraphics[width=\linewidth]{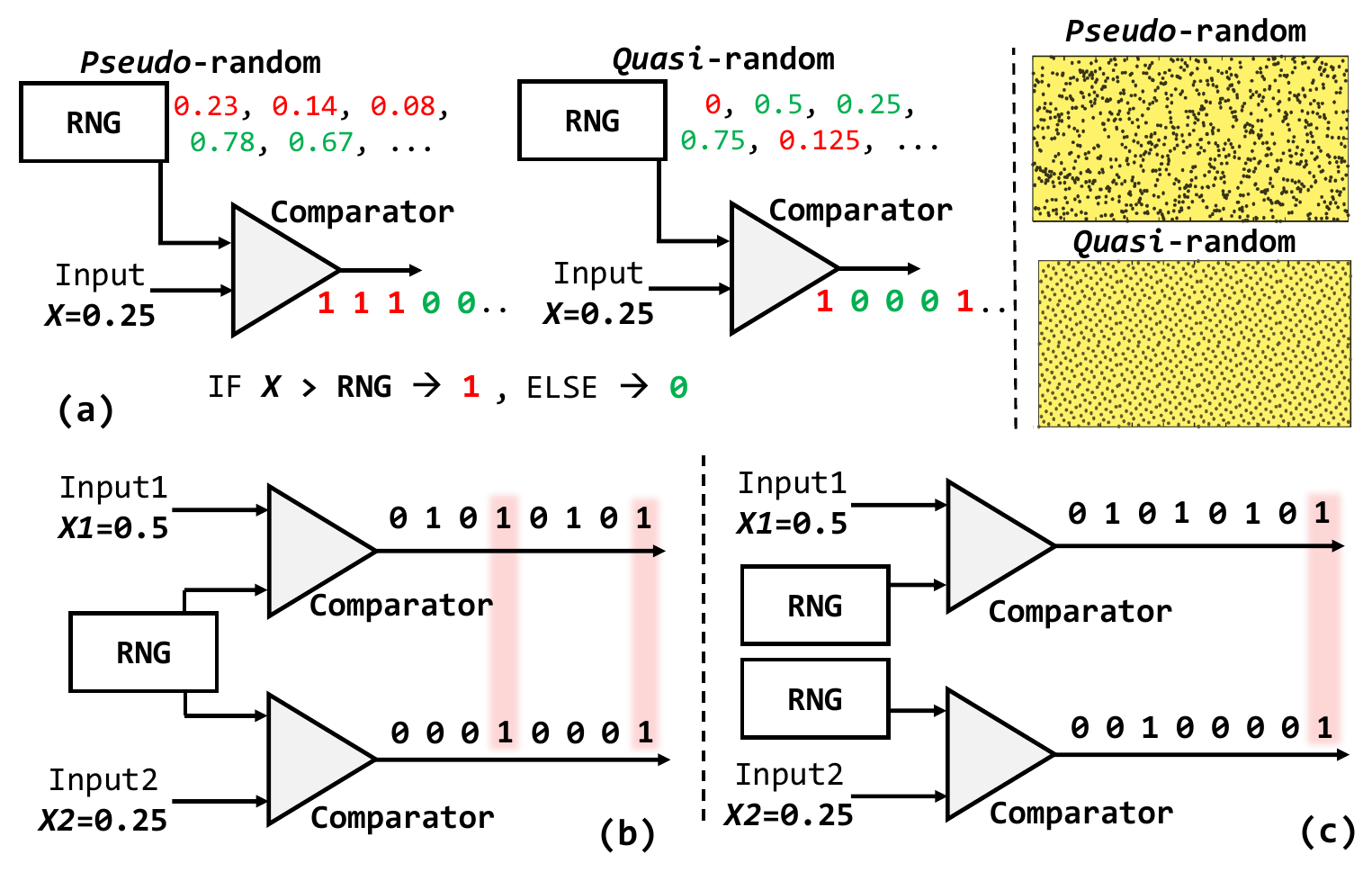}
        \vspace{-2em}
        \caption{Comparator-based SBS generation using pseudo-random and quasi-random sequences for probabilities $X_1$ and $X_2$.}
        \label{fig1} 
        \vspace{-.5em}
\end{figure}

The most common approach to generate SBS is \textit{comparator-based conversion}. In this approach, a sequence of $M$-bit  random binary numbers is produced using either a PRNG, such as an LFSR, or a QRNG, such as a Sobol or Halton sequence~\cite{9406123, liu2017energy, 10.5555/647234.719762}. Each random binary number %typically represented with $M$ bits, 
is compared with a target binary value $X_{bin}$ to produce one bit of the bit-stream~\cite{8782977}. As Fig. ~\ref{fig1}(a) shows, the output bit is one when the random number is less than $X_{bin}$ and zero otherwise. Repeating this over $N$ random numbers produces an SBS of length $N$ whose fraction of ones approximates the target probability $X$. This method underlies most SC systems due to its generality and ability to represent any value in $[0,1]$  to a resolution of $2^{-M}$~\cite{PengRiedelImageProcessingCaseStudies}.  
However, the hardware cost of  RNGs, particularly QRNGs, %that can achieve higher accuracy, 
can be significant, often dominating the cost of an SC design~\cite{8049760}. Furthermore, the reliance on deterministic PRNG or QRNG sequences can introduce unwanted correlation and predictability, which may compromise both accuracy and security.
%Fig.~\ref{fig1} illustrates the conventional comparator-based method for generating SBS. %: PRNGs and QRNGs produce sequences that are compared against a target probability $X$, outputting 1s when $X >$ RNG and 0s otherwise. Subfigure 
Fig. ~\ref{fig1}(a) illustrates the difference between pseudo-random and quasi-random sequences by plotting two sequences from the same source against each other. As can be seen, pseudo-random values scatter irregularly while quasi-random values fall into a regular pattern.
%samples form more structured patterns. %presents sample random-number sequence values and their distribution in the filling space (on the right), where the x- and y-axes represent two sequences from the same source type to show whether the sequences are related randomly or deterministically. 
Fig. s~\ref{fig1}(b) and (c) demonstrate two input values ($X_1 = 0.5$ and $X_2 = 0.25$)  converted into  \textit{correlated} bit-streams by sharing one RNG and into \textit{independent} bit-streams by using two RNGs.  
\section{Proposed True-Random Bit-Stream Generation Methods}
\label{prop}
%\vspace{-0.25em}
%In this section, we describe the three proposed true-random bit-stream generation methods, converting unbiased random bits into stochastic bit-streams with desired probabilities.
In this section, we discuss the three methods to %for generating true-random bit-streams. These approaches 
convert unbiased random bits into SBS with the desired probability.

\begin{figure}[t]
    \centering
        \includegraphics[width=0.8\linewidth]{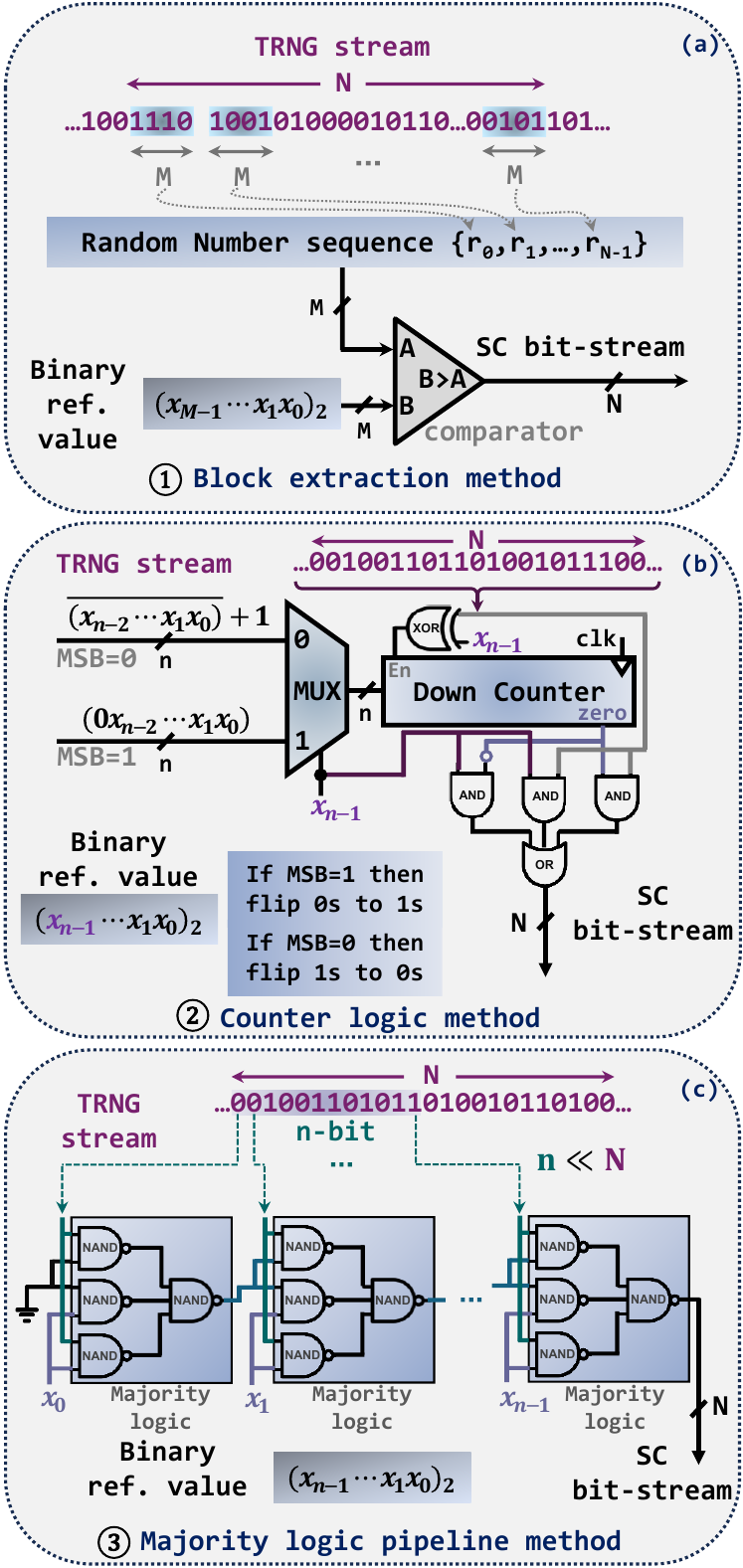}
        \vspace{-0.5em}
        \caption{\textit{From TRNG streams to SC bit-streams} with arbitrary probability values: (a) Block extraction method, (b) Counter logic method, and (c) Majority logic pipeline method.}
        \label{fig:bs_trng} 
        \vspace{-1.0em}
\end{figure}

\vspace{0.15em}
\textbf{\textit{Method 1. Block extraction \& comparison method.}}
The first %proposed 
method employs a \textit{block extraction \& comparison mechanism} inspired by conventional comparator-based bit-stream generation. In traditional SC designs, a random number is compared against a target value to decide each output bit (Fig. ~\ref{fig1}). We extend this principle by partitioning the unbiased TRNG output into fixed-length blocks of $M$ bits, each read as a uniformly distributed binary number in the range $[0,2^M-1]$. 
The target probability $X \in [0,1]$ is scaled to an $M$-bit binary value $X_{bin}$ (in conventional binary format). For each block, the output bit is one if the block value is less than $X_{bin}$ and zero otherwise. Repeating this process across $N$ blocks produces an SBS of length $N$ with the desired proportion of ones approximating $X$.
The method keeps the  \textit{generality} of comparator-based conversion and gains the \textit{unpredictability} of the source. %It is also \textit{scalable} and \textit{tunable}: 
Two parameters tune it: a larger block size $M$ gives finer resolution in $X$, and a longer stream $N$ reduces the variance of the encoded value.
%increasing the block size $M$ enhances the resolution of probability encoding,
%and increasing the stream length $N$ reduces variance. %\blue{by improving the bit-stream resolution.} 
Fig. ~\ref{fig:bs_trng}(a) shows the architecture. %for the proposed method. 
The required hardware is modest, consisting of an $M$-bit comparator and a register for the target value. %, making the design straightforward and CMOS-friendly.

\vspace{0.15em}
\textbf{\textit{Method 2. Counter-logic method.}}
The second %proposed 
method is built on a \textit{counter-logic architecture} that eliminates the need for %explicit 
multi-bit comparators by leveraging a down-counter and lightweight combinational logic. As shown in Fig. ~\ref{fig:bs_trng}(b), an unbiased TRNG stream provides the raw random bits, which are sequentially fed into the counter-logic block. The target probability value $X$ is stored as an $n$-bit binary value $X_{bin} = x_{n-1}\cdots x_0$ (binary format), and the converter operates on a window of $2^n$ source bits, which on average contain $2^{n-1}$ ones. The most significant bit (MSB) $x_{n-1}$ decides the direction of the flips. If $x_{n-1}=1$, then $X \geq 0.5$ and the window needs more ones, so the converter flips the first $L = (x_{n-2}\cdots x_0)_2$ zeros it encounters. If $x_{n-1}=0$, then $X < 0.5$, so the converter flips the first $2^{n-1}-L$ ones to zeros. The two flip counts are the lower $n-1$ bits of $X_{bin}$ and their two's complement, and a \texttt{MUX} controlled by the MSB loads the selected count into the down-counter. An \texttt{XOR} of the incoming bit with the MSB flags each flippable bit (a zero when $x_{n-1}=1$, a one when $x_{n-1}=0$) and enables the counter. While the counter is nonzero, the output logic forces the flip. When the counter reaches zero it raises a flag, and the remaining source bits pass through unchanged. The expected number of ones in the window is $2^{n-1}+L$ when $x_{n-1}=1$ and $L$ when $x_{n-1}=0$. Both equal $X \cdot 2^{n}$.  
%determines whether the design flips zeros to ones or ones to zeros in the generated output stream, ensuring correct probability adjustment. Specifically, if $\text{MSB}=1$, the logic inverts a subset of zeros to ones; if $\text{MSB}=0$, the logic inverts a subset of ones to zeros.   The down-counter monitors the incoming TRNG bits and asserts a zero flag when the number of required flips has been satisfied. A \texttt{MUX} controlled by the MSB selects between the two input paths: one corresponding to $X_{bin}$ and the other to its two’s complement $((\sim X_{bin})+1)$ (where $\sim$ is bitwise inversion). This structure ensures that the probability of ones in the output bit-stream converges to the target value $X$.  

%By replacing costly comparators with a simple counter, multiplexer, and a few logic gates, this method provides a \textit{hardware-efficient alternative} for TRNG-based stochastic bit-stream generation. It retains accuracy while reducing implementation overhead, making it well-suited for compact, low-power CMOS designs.

\vspace{0.15em}
\textbf{\textit{Method 3. Majority-logic pipeline method.}} The third method employs a \textit{majority-logic pipeline architecture} that uses cascaded majority gates to directly map unbiased TRNG bits into SBS with the desired probability. Unlike comparator- or counter-based designs, which require multi-bit comparisons or sequential updates, this method embeds the probability transformation into a %purely 
combinational structure of majority gates. %, enabling high throughput and scalability. 
Fig. ~\ref{fig:bs_trng}(c) shows the circuit.
The process begins by storing the target probability $X \in [0,1]$ as an $n$-bit binary reference 
value $X_{bin}$. %(\blue{conventional binary representation}). 
The converter is a chain of $n$ three-input majority gates, each built from four \texttt{NAND} gates.
Stage $i$ receives the output of stage $i-1$, one fresh source bit $r_i$, and the reference bit $x_i$. The chain input of the first stage is tied to zero. With a constant reference bit, a majority gate reduces to a two-input function: $\mathrm{MAJ}(p,r,1) = p \lor r$ and $\mathrm{MAJ}(p,r,0) = p \land r$.  Since $r_i$ is one with probability $1/2$ and independent of the chain input, each stage maps the probability of a one from $p_i$ to $(p_i + x_i)/2$. Starting from $p_0 = 0$ and applying the stages from the least to the MSB gives 
\vspace{-0.5em}
\begin{equation} p_n = \sum_{i=0}^{n-1} x_i\, 2^{\,i-n} = X. \end{equation} 
Each output bit consumes $n$ source bits, one per stage.
%Each bit of $X_{bin}$ is paired with a segment of TRNG output bits, and the pair is fed into a stage of majority logic. For a given stage, the majority gate decides whether more inputs favor a ‘1’ or a ‘0’, thereby biasing the output toward the desired probability. Cascading multiple stages ensures that the statistical contribution of each reference bit is accumulated across the pipeline. After passing through all $n$ stages, the output converges to a stochastic bit-stream whose fraction of ones closely approximates $X$.  
The pipeline has three properties. 
%First, it is \textit{structurally simple and small}, as the majority gates are low-cost primitives.
It is small: $n$ majority gates and no other logic. 
It pipelines: with a register between stages, the chain accepts $n$ fresh source bits every cycle and delivers one output bit per cycle. 
%Second, the pipeline arrangement supports \textit{parallelism and high throughput}, as multiple TRNG bits are processed in parallel. 
And its precision is set by $n$ alone: each added stage halves the resolution step in $X$ at the cost of one gate and one source bit per output.

%The third method offers three key advantages. First, it is \textit{structurally simple}, as the majority gates are low-cost, CMOS-friendly primitives. Second, the pipeline arrangement supports \textit{parallelism and high throughput}, as multiple TRNG bits are processed in parallel.Finally, it provides a \textit{flexible trade-off between accuracy and cost}: increasing the number of stages $n$ improves resolution and reduces variance in the generated stream.  As shown in Figure~\ref{fig:bs_trng}(c), this design generates SBS without explicit comparators or counters, making it highly suitable for lightweight, low-power SC systems.

{\textbf{\textit{Correlation Control.}}
In SC systems, both correlated and uncorrelated bit-streams are needed, depending on the desired function. 
%Correct functionality of SC operations depends on the level of correlation between input bit-streams. 
Some operations, such as multiplication using bit-wise \texttt{AND}, require \textit{independent} (i.e., uncorrelated) inputs, while some others, such as absolute value subtraction using bit-wise \texttt{XOR}, require \textit{maximal correlation}~\cite{AygunAgile}.} 
The proposed converters support both. 
%the three proposed methods differ in how they handle \textit{correlation} between generated streams. 
The \textit{block-extraction} method can produce \textit{independent} bit-streams when distinct blocks of one or multiple TRNG sequences are used to convert the input binary values, while sharing similar blocks naturally introduces \textit{correlation}. The \textit{counter-logic} and \textit{majority-logic pipeline}  converters behave the same way;
%inherit a similar property: 
independence is guaranteed only if separate random sequences are provided to each converter instance, whereas reusing the same TRNG source leads to (partially) \textit{correlated} outputs. {We will evaluate the performance of the proposed methods in generating correlated and independent bit-streams in Section~\ref{perf}\textcolor{blue}{.3.}}

\begin{table}[!t]
\centering
\caption{MAE (\%) Comparison of SC Bit-stream Generation Methods}
\vspace{-.75em}
\setlength{\tabcolsep}{2pt}
%\renewcommand{\arraystretch}{0.5}
% Use the D column type for decimal alignment
%\begin{tabular}{l l *{8}{D{.}{.}{2.3}}}
\resizebox{\columnwidth}{!}{
\begin{tabular}{l c *{8}{c}}
\toprule
\multirow{2}{*}{\textbf{SBS Generation Method}} & \multirow{2}{*}{\begin{tabular}[c]{@{}c@{}}\textbf{Block}\\\textbf{size}\textbf{(M)} \end{tabular}} & \multicolumn{8}{c}{\textbf{Bit-stream length (N)}} \\
\cmidrule(lr){3-10}
% Wrap each header number in \multicolumn to center it
& & \multicolumn{1}{c}{\textbf{8}} & \multicolumn{1}{c}{\textbf{16}} & \multicolumn{1}{c}{\textbf{32}} & \multicolumn{1}{c}{\textbf{64}} & \multicolumn{1}{c}{\textbf{128}} & \multicolumn{1}{c}{\textbf{256}} & \multicolumn{1}{c}{\textbf{512}} & \multicolumn{1}{c}{\textbf{1024}} \\
\midrule

\multirow{8}{*}{\begin{tabular}[c]{@{}c@{}} %\textbf{Proposed Method \ding{172}} \\ 
\textbf{\textit{\ding{172}~Block Extraction}}\\+\textbf{\textcolor{mysecpurple}{TRNG1}}~\cite{TRNG_randomorg} \end{tabular}} &\cellcolor{Lavender!40}3 &\cellcolor{Lavender!40}12.5 &\cellcolor{Lavender!40}9.89 &\cellcolor{Lavender!40}8.21 &\cellcolor{Lavender!40}7.27 &\cellcolor{Lavender!40}6.77 &\cellcolor{Lavender!40}6.52 &\cellcolor{Lavender!40}6.39 &\cellcolor{Lavender!40}6.30 \\
%\rowcolor[HTML]{EFEFEF}
& 4 & 11.6 & 8.40 & 6.33 & 4.98 & 4.11 & 3.63 & 3.40 & 3.25 \\
&\cellcolor{Lavender!40}5 &\cellcolor{Lavender!40}11.3 &\cellcolor{Lavender!40}8.01 &\cellcolor{Lavender!40}5.78 &\cellcolor{Lavender!40}4.20 &\cellcolor{Lavender!40}3.17 &\cellcolor{Lavender!40}2.48 &\cellcolor{Lavender!40}2.06 &\cellcolor{Lavender!40}1.83 \\
%\rowcolor[HTML]{EFEFEF}
& 6 & 11.2 & 7.93 & 5.62 & 4.00 & 2.90 & 2.10 & 1.58 & 1.24 \\
&\cellcolor{Lavender!40}7 &\cellcolor{Lavender!40}11.1 &\cellcolor{Lavender!40}7.91 &\cellcolor{Lavender!40}5.54 &\cellcolor{Lavender!40}3.93 &\cellcolor{Lavender!40}2.79 &\cellcolor{Lavender!40}1.99 &\cellcolor{Lavender!40}1.43 &\cellcolor{Lavender!40}1.05 \\
%\rowcolor[HTML]{EFEFEF}
& 8 & 11.1 & 7.89 & 5.55 & 3.93 & 2.79 & 1.97 & 1.40 & 1.00 \\
&\cellcolor{Lavender!40}9 &\cellcolor{Lavender!40}11.1 &\cellcolor{Lavender!40}7.87 &\cellcolor{Lavender!40}5.54 &\cellcolor{Lavender!40}3.92 &\cellcolor{Lavender!40}2.77 &\cellcolor{Lavender!40}1.96 &\cellcolor{Lavender!40}1.39 &\cellcolor{Lavender!40}0.99 \\
%\rowcolor[HTML]{EFEFEF}
& 10 & 11.1 & 7.85 & 5.55 & 3.93 & 2.78 & 1.96 & 1.38 & 0.97 \\
%\addlinespace % Adds a little extra vertical space for separation
\midrule
\multirow{8}{*}{\begin{tabular}[c]{@{}c@{}} %\textbf{Proposed Method \ding{172}} \\
\textbf{\textit{\ding{172}~Block Extraction}}\\+\textbf{\textcolor{mysecpurple}{TRNG2}}~\cite{TRNG_randombeacons} \end{tabular}} &\cellcolor{Lavender!40}3 &\cellcolor{Lavender!40}12.7 &\cellcolor{Lavender!40}10.0 &\cellcolor{Lavender!40}8.17 &\cellcolor{Lavender!40}7.20 &\cellcolor{Lavender!40}6.82 &\cellcolor{Lavender!40}6.49 &\cellcolor{Lavender!40}6.38 &\cellcolor{Lavender!40}6.32 \\
& 4 & 11.6 & 8.37 & 6.32 & 4.91 & 4.13 & 3.65 & 3.38 & 3.26 \\
&\cellcolor{Lavender!40}5 &\cellcolor{Lavender!40}11.1 &\cellcolor{Lavender!40}8.04 &\cellcolor{Lavender!40}5.68 &\cellcolor{Lavender!40}4.23 &\cellcolor{Lavender!40}3.18 &\cellcolor{Lavender!40}2.49 &\cellcolor{Lavender!40}2.05 &\cellcolor{Lavender!40}1.81 \\
& 6 & 11.3 & 7.93 & 5.61 & 3.99 & 2.89 & 2.11 & 1.57 & 1.22 \\
&\cellcolor{Lavender!40}7 &\cellcolor{Lavender!40}11.2 &\cellcolor{Lavender!40}7.95 &\cellcolor{Lavender!40}5.56 &\cellcolor{Lavender!40}3.99 &\cellcolor{Lavender!40}2.78 &\cellcolor{Lavender!40}2.01 &\cellcolor{Lavender!40}1.45 &\cellcolor{Lavender!40}1.06 \\
& 8 & 11.2 & 7.82 & 5.53 & 3.85 & 2.76 & 1.96 & 1.37 & 0.97 \\
&\cellcolor{Lavender!40}9 &\cellcolor{Lavender!40}11.1 &\cellcolor{Lavender!40}8.06 &\cellcolor{Lavender!40}5.50 &\cellcolor{Lavender!40}3.99 &\cellcolor{Lavender!40}2.76 &\cellcolor{Lavender!40}1.96 &\cellcolor{Lavender!40}1.37 &\cellcolor{Lavender!40}0.95 \\
& 10 & 11.2 & 7.77 & 5.59 & 3.92 & 2.79 & 1.96 & 1.38 & 0.97 \\
\midrule

% Other data rows
\begin{tabular}[c]{@{}c@{}}  %\textbf{Proposed Method \ding{174}}\\ 
\textbf{\textit{\ding{173}~Counter Logic}+\textcolor{mysecpurple}{TRNG1}}~\cite{TRNG_randomorg} \end{tabular} & \cellcolor{Lavender!40}{--} &\cellcolor{Lavender!40}12.1 &\cellcolor{Lavender!40}9.01 &\cellcolor{Lavender!40}6.59 &\cellcolor{Lavender!40}4.74 &\cellcolor{Lavender!40}3.40 &\cellcolor{Lavender!40}2.43 &\cellcolor{Lavender!40}1.74 &\cellcolor{Lavender!40}1.24 \\
%\midrule
\textbf{\textit{\ding{173}~Counter Logic}+\textcolor{mysecpurple}{TRNG2}}~\cite{TRNG_randombeacons} & {--} & 12.0 & 9.00 & 6.63 & 4.80 & 3.46 & 2.46 & 1.74 & 1.22 \\
%\addlinespace
\midrule
\textbf{\textit{\ding{174}~Majority Logic}+\textcolor{mysecpurple}{TRNG1}}~\cite{TRNG_randomorg} & \cellcolor{Lavender!40}{--} &\cellcolor{Lavender!40}22.5 &\cellcolor{Lavender!40}13.7 &\cellcolor{Lavender!40}8.57 &\cellcolor{Lavender!40}5.41 &\cellcolor{Lavender!40}3.57 &\cellcolor{Lavender!40}2.35 &\cellcolor{Lavender!40}1.58 &\cellcolor{Lavender!40}1.06 \\
\textbf{\textit{\ding{174}~Majority Logic}+\textcolor{mysecpurple}{TRNG2}}~\cite{TRNG_randombeacons} & {--} & 22.6 & 13.7 & 8.57 & 5.50 & 3.52 & 2.29 & 1.57 & 1.06 \\
\midrule

% Final data rows
\texttt{MATLAB} (rand) & \cellcolor{Lavender!40}{--} &\cellcolor{Lavender!40}11.2 &\cellcolor{Lavender!40}7.84 &\cellcolor{Lavender!40}5.54 &\cellcolor{Lavender!40}3.91 &\cellcolor{Lavender!40}2.76 &\cellcolor{Lavender!40}1.96 &\cellcolor{Lavender!40}1.38 &\cellcolor{Lavender!40}0.97 \\
PRNG ($\log_2{N}$-bit LFSR) & {--} & 11.8 & 6.10 & 3.17 & 1.51 & 0.77 & 0.39 & 0.21 & 0.10 \\
QRNG (Sobol) & \cellcolor{Lavender!40}{--} &\cellcolor{Lavender!40}6.25 &\cellcolor{Lavender!40}3.13 &\cellcolor{Lavender!40}1.56 &\cellcolor{Lavender!40}0.78 &\cellcolor{Lavender!40}0.39 &\cellcolor{Lavender!40}0.19 &\cellcolor{Lavender!40}0.09 &\cellcolor{Lavender!40}0.04 \\
\bottomrule
%\multicolumn{10}{l}{\scriptsize For PRNG, we use a maximal length LFSR with polynomial $x^8+x^5+x^3+1$.} % Table note, left-aligned
\end{tabular}
}
\label{tab:accuracy_bs}
\vspace{-1.em}
\end{table}

\section{Performance Evaluation}
\label{perf}
This section evaluates the proposed methods on four fronts: the accuracy of the generated SBS, the accuracy of SC arithmetic performed on them, the correlation between streams generated together, and their performance in an image-processing case study.
%{In this section, we evaluate the performance of the proposed methods in generating SBS, performing SC arithmetic operations on those SBS, and the correlation %/independence level between the generated SBS.} 

\subsection{Stochastic Bit-stream Generation}
%{\textbf{\textit{4.1. SBS Generation:}}} %In this section, 
We first evaluate the accuracy of the three converters %in generating SBS against 
with SOTA PRNG- and QRNG-based generators and with a software generator that uses \texttt{MATLAB}'s \texttt{rand} function as the RNG. For each method, we convert 100{,}000 random real numbers in the $[0,1]$ interval into SBS of lengths $N \in \{8, 16, 32, 64, 128, 256, 512, 1024\}$ and measure the mean absolute error (MAE) between each target value and the value the stream encodes. Table~\ref{tab:accuracy_bs} reports the results.
For optimal performance of the PRNG baseline, we implement maximal-period LFSRs sized according to the bit-stream length ($N$), i.e., a $\log_2{N}$-bit maximal-period LFSR is used to generate an $N$-bit bit-stream. %a polynomial feedback function of $x^8+x^5+x^3+1$. 
For the QRNG baseline, we use the first Sobol sequence from \texttt{MATLAB}'s %built-in
Sobol source generator~\cite{AygunOptimization, 8327916}. 
%For the three proposed methods, we input true random sequences produced by two true-random input sources 
%For the true random inputs to the proposed methods of this work, we use two sequences of random bits 
%from
The three converters draw their input bits from two true-random sources, TRNG1~\cite{TRNG_randomorg}  and TRNG2~\cite{TRNG_randombeacons}, both of which have successfully passed all NIST statistical random tests~\cite{nistspecialpublication800-22r1a}.

As shown in Table~\ref{tab:accuracy_bs}, all three proposed TRNG-to-SBS converters reach accuracy comparable to the software-based \texttt{MATLAB} generator, which confirms that they turn unbiased
%their effectiveness in converting unbiased 
random bits into well-formed  SBS. The block-extraction and counter-logic methods achieve particularly low MAE values across different bit-stream lengths, and the majority-logic method converges to \texttt{MATLAB}-level accuracy by $N$=$256$.
%maintains accuracy at a level very close to MATLAB. 
The LFSR and Sobol baselines report lower MAE than any true-random source, including MATLAB. This is expected: a maximal-period LFSR sized to $N$ visits each value exactly once per stream, and a Sobol sequence is low-discrepancy by construction, so both remove the sampling variance that a random source carries. That property is exactly what makes them predictable and correlated across streams~\cite{Heese_2024}.
%Although the MAEs of the proposed methods are generally higher than those achieved by PRNG- and QRNG-based generators, 
%The key distinguishing advantage lies in their reliance on true randomness rather than deterministic sequences. This intrinsic unpredictability directly addresses the limitations of PRNG and QRNG approaches~\cite{Heese_2024}, offering enhanced robustness, security, and resilience; benefits that outweigh the marginal differences in accuracy. 
The proposed converters therefore sit at the accuracy of true randomness, which is the floor for any unpredictable source, and trade the residual variance for streams that no observer can reproduce. 
%Moreover, the results for both TRNG1 and TRNG2 are very close across all three methods, confirming that the accuracy of each design is \textit{largely independent of the specific true-random input source} and can be expected to generalize to other TRNGs as well.
Results for TRNG1 and TRNG2 agree to within 0.2\% for all three methods, so the accuracy of each converter does not depend on which true-random source feeds it.
%Collectively, these results confirm that TRNG-to-SBS conversion using CMOS logic is a practical and competitive solution for SC.

\begin{table}[t]
\centering
\caption{MAE (\%) Comparison of SC Arithmetic Operations using different SBS Generation Methods}
% Set up the S column type for numbers with max 2 digits before and 2 after the decimal.
%\sisetup{table-format=2.2}
\vspace{-.75em}
\setlength{\tabcolsep}{2pt}
\renewcommand{\arraystretch}{0.95}
\resizebox{\columnwidth}{!}{
\begin{tabular}{ll *{8}{c}}
\toprule
% Table Header
\multirow{2}{*}{\begin{tabular}[c]{@{}l@{}}\textbf{SBS} \textbf{Gen.}\\ \textbf{Method} \end{tabular}} & \multirow{2}{*}{\textbf{\begin{tabular}[c]{@{}l@{}}SC Arithmetic \\ Operation\end{tabular}}} & \multicolumn{8}{c}{\textbf{Bit-stream length (N)}} \\
\cmidrule(lr){3-10} % Partial rule under the main header
& & \textbf{8} & \textbf{16} & \textbf{32} & \textbf{64} & \textbf{128} & \textbf{256} & \textbf{512} & \textbf{1024} \\
\midrule

\multirow{7}{*}{\begin{tabular}[c]{@{}l@{}}\textbf{\ding{172} \textit{Block}} \\ \textbf{\textit{Extraction}} \\ +\textbf{\textcolor{mysecpurple}{TRNG1}}\cite{TRNG_randomorg} \\ (M=8)\end{tabular}}
&\cellcolor{Lavender!40}Multiplication&\cellcolor{Lavender!40}9.86 &\cellcolor{Lavender!40}7.01 &\cellcolor{Lavender!40}4.95 &\cellcolor{Lavender!40}3.48 &\cellcolor{Lavender!40}2.47 &\cellcolor{Lavender!40}1.76 &\cellcolor{Lavender!40}1.26 &\cellcolor{Lavender!40}0.89 \\
& Scaled Addition & 13.1 & 8.99 & 6.47 & 4.49 & 3.19 & 2.27 & 1.61 & 1.13 \\
%&  Approx. Add  & 14.2 & 9.77 & 6.96 & 4.88 & 6.90 & 6.68 & 6.64 & 6.71 \\
&\cellcolor{Lavender!40}Abs. Subtraction &\cellcolor{Lavender!40}11.2 &\cellcolor{Lavender!40}7.86 &\cellcolor{Lavender!40}5.57 &\cellcolor{Lavender!40}3.91 &\cellcolor{Lavender!40}2.77 &\cellcolor{Lavender!40}1.98 &\cellcolor{Lavender!40}1.40 &\cellcolor{Lavender!40}1.00 \\
& Division & 16.8 & 12.1 & 8.69 & 6.08 & 4.39 & 3.11 & 2.18 & 1.59 \\
&\cellcolor{Lavender!40}Minimum &\cellcolor{Lavender!40}11.3 &\cellcolor{Lavender!40}7.87 &\cellcolor{Lavender!40}5.62 &\cellcolor{Lavender!40}3.91 &\cellcolor{Lavender!40}2.80 &\cellcolor{Lavender!40}2.03 &\cellcolor{Lavender!40}1.40 &\cellcolor{Lavender!40}1.01 \\
& Maximum & 11.2 & 7.70 & 5.58 & 3.88 & 2.77 & 1.96 & 1.40 & 1.01 \\
\midrule

% Data Section 4
\multirow{7}{*}{\begin{tabular}[c]{@{}l@{}}\textbf{\ding{172} \textit{Block}} \\ \textbf{\textit{Extraction}} \\ +\textbf{\textcolor{mysecpurple}{TRNG2}}\cite{TRNG_randombeacons} \\ (M=8)\end{tabular}}
&\cellcolor{Lavender!40}Multiplication&\cellcolor{Lavender!40}9.99 &\cellcolor{Lavender!40}6.93 &\cellcolor{Lavender!40}4.86 &\cellcolor{Lavender!40}3.46 &\cellcolor{Lavender!40}2.45 &\cellcolor{Lavender!40}1.74 &\cellcolor{Lavender!40}1.23 &\cellcolor{Lavender!40}0.87 \\
& Scaled Addition & 12.8 & 9.06 & 6.33 & 4.47 & 3.13 & 2.23 & 1.57 & 1.12 \\
%& Approx. Add  & 14.2 & 11.10 & 8.09 & 7.06 & 7.06 & 6.81 & 6.24 & 6.42 \\
&\cellcolor{Lavender!40}Abs. Subtraction &\cellcolor{Lavender!40}11.2 &\cellcolor{Lavender!40}7.79 &\cellcolor{Lavender!40}5.45 &\cellcolor{Lavender!40}3.92 &\cellcolor{Lavender!40}2.71 &\cellcolor{Lavender!40}1.91 &\cellcolor{Lavender!40}1.34 &\cellcolor{Lavender!40}0.93 \\
& Division & 16.7 & 12.07 & 8.61 & 6.02 & 4.29 & 3.06 & 2.19 & 1.56 \\
& \cellcolor{Lavender!40}Minimum &\cellcolor{Lavender!40}11.2 &\cellcolor{Lavender!40}7.80 &\cellcolor{Lavender!40}5.49 &\cellcolor{Lavender!40}3.84 &\cellcolor{Lavender!40}2.72 &\cellcolor{Lavender!40}1.91 &\cellcolor{Lavender!40}1.36 &\cellcolor{Lavender!40}0.96 \\
& Maximum & 11.2 & 7.78 & 5.47 & 3.86 & 2.70 & 1.86 & 1.32 & 0.95 \\
\midrule

% Data Section 1
\multirow{7}{*}{\begin{tabular}[c]{@{}l@{}} \textbf{\ding{173} \textit{Counter}}\\\textbf{\textit{Logic}}\\ +\textbf{\textcolor{mysecpurple}{TRNG1}}\cite{TRNG_randomorg}\end{tabular}}
&\cellcolor{Lavender!40}Multiplication &\cellcolor{Lavender!40}12.2 &\cellcolor{Lavender!40}8.28 &\cellcolor{Lavender!40}5.85 &\cellcolor{Lavender!40}4.13 &\cellcolor{Lavender!40}2.94 &\cellcolor{Lavender!40}2.07 &\cellcolor{Lavender!40}1.47 &\cellcolor{Lavender!40}1.04 \\
& Scaled Addition & 15.8 & 11.3 & 8.02 & 5.68 & 4.08 & 2.88 & 2.06 & 1.46 \\
%& Approx. Add& 16.2 & 12.0 & 8.76 & 6.17 & 4.38 & 2.93 & 2.01 & 1.40 \\
&\cellcolor{Lavender!40}Abs. Subtraction &\cellcolor{Lavender!40}12.3 &\cellcolor{Lavender!40}8.30 &\cellcolor{Lavender!40}6.32 &\cellcolor{Lavender!40}5.11 &\cellcolor{Lavender!40}4.72 &\cellcolor{Lavender!40}4.58 &\cellcolor{Lavender!40}4.29 &\cellcolor{Lavender!40}4.18 \\
& Division& 18.7 & 14.1 & 10.8 & 9.06 & 7.82 & 7.23 & 6.40 & 6.18 \\
&\cellcolor{Lavender!40}Minimum&\cellcolor{Lavender!40}11.5 &\cellcolor{Lavender!40}8.80 &\cellcolor{Lavender!40}6.52 &\cellcolor{Lavender!40}4.75 &\cellcolor{Lavender!40}3.40 &\cellcolor{Lavender!40}2.45 &\cellcolor{Lavender!40}1.74 &\cellcolor{Lavender!40}1.23 \\
& Maximum & 12.8 & 9.31 & 6.71 & 4.82 & 3.43 & 2.46 & 1.74 & 1.24 \\
\midrule

% Data Section 2
\multirow{7}{*}{\begin{tabular}[c]{@{}l@{}} \textbf{\ding{173} \textit{Counter}}\\\textbf{\textit{Logic}} \\ +\textbf{\textcolor{mysecpurple}{TRNG2}}\cite{TRNG_randombeacons}\end{tabular}}
&\cellcolor{Lavender!40}Multiplication&\cellcolor{Lavender!40}12.2 &\cellcolor{Lavender!40}8.33 &\cellcolor{Lavender!40}5.90 &\cellcolor{Lavender!40}4.10 &\cellcolor{Lavender!40}2.91 &\cellcolor{Lavender!40}2.10 &\cellcolor{Lavender!40}1.48 &\cellcolor{Lavender!40}1.05 \\
& Scaled Addition & 16.0 & 11.3 & 8.14 & 5.74 & 4.08 & 2.91 & 2.05 & 1.47 \\
%& Approx. Add& 16.0 & 11.8 & 8.61 & 6.20 & 4.40 & 3.01 & 2.06 & 1.46 \\
&\cellcolor{Lavender!40}Abs. Subtraction &\cellcolor{Lavender!40}12.4 &\cellcolor{Lavender!40}8.23 &\cellcolor{Lavender!40}6.33 &\cellcolor{Lavender!40}5.31 &\cellcolor{Lavender!40}4.85 &\cellcolor{Lavender!40}4.41 &\cellcolor{Lavender!40}4.34 &\cellcolor{Lavender!40}4.26 \\
& Division& 18.9 & 14.0 & 11.2 & 9.42 & 8.11 & 7.08 & 6.59 & 6.24 \\
&\cellcolor{Lavender!40}Minimum&\cellcolor{Lavender!40}11.4 &\cellcolor{Lavender!40}8.78 &\cellcolor{Lavender!40}6.51 &\cellcolor{Lavender!40}4.78 &\cellcolor{Lavender!40}3.45 &\cellcolor{Lavender!40}2.47 &\cellcolor{Lavender!40}1.74 &\cellcolor{Lavender!40}1.22 \\
& Maximum& 12.7 & 9.29 & 6.71 & 4.86 & 3.48 & 2.48 & 1.74 & 1.22 \\
\midrule

\multirow{7}{*}{\begin{tabular}[c]{@{}l@{}}\textbf{\ding{174} \textit{Majority}}\\\textbf{\textit{Logic}} \\ +\textbf{\textcolor{mysecpurple}{TRNG1}}\cite{TRNG_randomorg}\end{tabular}}
&\cellcolor{Lavender!40}Multiplication &\cellcolor{Lavender!40}20.4 &\cellcolor{Lavender!40}13.1 &\cellcolor{Lavender!40}7.85 &\cellcolor{Lavender!40}5.04 &\cellcolor{Lavender!40}3.15 &\cellcolor{Lavender!40}2.24 &\cellcolor{Lavender!40}1.49 &\cellcolor{Lavender!40}0.95 \\
& Scaled Addition & 22.4 & 14.4 & 9.11 & 6.02 & 3.87 & 2.70 & 1.87 & 1.21 \\
%& Approx. Add & 22.4 & 14.3 & 9.21 & 7.82 & 7.18 & 6.69 & 6.52 & 6.31 \\
&\cellcolor{Lavender!40}Abs. Subtraction &\cellcolor{Lavender!40}17.0 &\cellcolor{Lavender!40}12.3 &\cellcolor{Lavender!40}7.97 &\cellcolor{Lavender!40}5.49 &\cellcolor{Lavender!40}3.40 &\cellcolor{Lavender!40}2.36 &\cellcolor{Lavender!40}1.64 &\cellcolor{Lavender!40}1.05 \\
& Division & 26.4 & 17.8 & 11.32 & 7.74 & 5.00 & 3.56 & 2.47 & 1.60 \\
&\cellcolor{Lavender!40}Minimum &\cellcolor{Lavender!40}21.1 &\cellcolor{Lavender!40}13.6 &\cellcolor{Lavender!40}8.31 &\cellcolor{Lavender!40}5.49 &\cellcolor{Lavender!40}3.45 &\cellcolor{Lavender!40}2.45 &\cellcolor{Lavender!40}1.65 &\cellcolor{Lavender!40}1.05 \\
& Maximum & 21.2 & 13.4 & 8.29 & 5.50 & 3.43 & 2.33 & 1.69 & 1.06 \\
\midrule

% Data Section 6
\multirow{7}{*}{\begin{tabular}[c]{@{}l@{}}\textbf{\ding{174} \textit{Majority}}\\\textbf{\textit{Logic}} \\ +\textbf{\textcolor{mysecpurple}{TRNG2}}\cite{TRNG_randombeacons}\end{tabular}}
&\cellcolor{Lavender!40}Multiplication &\cellcolor{Lavender!40}21.1 &\cellcolor{Lavender!40}12.4 &\cellcolor{Lavender!40}7.97 &\cellcolor{Lavender!40}4.89 &\cellcolor{Lavender!40}3.24 &\cellcolor{Lavender!40}2.08 &\cellcolor{Lavender!40}1.40 &\cellcolor{Lavender!40}0.97 \\
& Scaled Addition & 22.7 & 14.3 & 9.11 & 5.99 & 3.91 & 2.56 & 1.75 & 1.23 \\
%& Approx. Add& 14.0 & 11.3 & 9.25 & 7.98 & 7.32 & 6.85 & 6.54 & 6.50 \\
&\cellcolor{Lavender!40}Abs. Subtraction &\cellcolor{Lavender!40}17.5 &\cellcolor{Lavender!40}12.0 &\cellcolor{Lavender!40}8.08 &\cellcolor{Lavender!40}5.30 &\cellcolor{Lavender!40}3.51 &\cellcolor{Lavender!40}2.30 &\cellcolor{Lavender!40}1.53 &\cellcolor{Lavender!40}1.08 \\
& Division & 26.9 & 17.3 & 11.4 & 7.55 & 5.12 & 3.43 & 2.28 & 1.63 \\
&\cellcolor{Lavender!40}Minimum&\cellcolor{Lavender!40}21.7 &\cellcolor{Lavender!40}13.1 &\cellcolor{Lavender!40}8.49 &\cellcolor{Lavender!40}5.38 &\cellcolor{Lavender!40}3.53 &\cellcolor{Lavender!40}2.29 &\cellcolor{Lavender!40}1.52 &\cellcolor{Lavender!40}1.06 \\
& Maximum & 21.2 & 13.4 & 8.25 & 5.53 & 3.47 & 2.22 & 1.57 & 1.09 \\
\midrule

% Data Section 7
\multirow{7}{*}{\begin{tabular}[c]{@{}l@{}} \texttt{MATLAB} \\ (rand)\end{tabular}}
&\cellcolor{Lavender!40}Multiplication &\cellcolor{Lavender!40}9.97 &\cellcolor{Lavender!40}6.97 &\cellcolor{Lavender!40}4.90 &\cellcolor{Lavender!40}3.50 &\cellcolor{Lavender!40}2.43 &\cellcolor{Lavender!40}1.72 &\cellcolor{Lavender!40}1.23 &\cellcolor{Lavender!40}0.86 \\
& Scaled Addition & 12.9 & 9.05 & 6.38 & 4.51 & 3.18 & 2.23 & 1.60 & 1.13 \\
%& Approx Add & 14.8 & 11.5 & 9.19 & 8.05 & 7.25 & 6.83 & 6.48 & 6.40 \\
&\cellcolor{Lavender!40}Abs. Subtraction &\cellcolor{Lavender!40}11.2 &\cellcolor{Lavender!40}7.75 &\cellcolor{Lavender!40}5.47 &\cellcolor{Lavender!40}3.92 &\cellcolor{Lavender!40}2.77 &\cellcolor{Lavender!40}1.97 &\cellcolor{Lavender!40}1.38 &\cellcolor{Lavender!40}0.98 \\
& Division & 16.7 & 12.1 & 8.59 & 6.17 & 4.37 & 3.12 & 2.17 & 1.55 \\
&\cellcolor{Lavender!40}Minimum &\cellcolor{Lavender!40}11.2 &\cellcolor{Lavender!40}7.89 &\cellcolor{Lavender!40}5.54 &\cellcolor{Lavender!40}3.92 &\cellcolor{Lavender!40}2.78 &\cellcolor{Lavender!40}1.94 &\cellcolor{Lavender!40}1.37 &\cellcolor{Lavender!40}0.99 \\
& Maximum & 11.1 & 7.94 & 5.52 & 3.90 & 2.74 & 1.96 & 1.38 & 0.98 \\
\midrule

% Data Section 8
\multirow{7}{*}{\begin{tabular}[c]{@{}l@{}}PRNG\\ ($\log_2{N}$-bit LFSR) \end{tabular}}
&\cellcolor{Lavender!40}Multiplication &\cellcolor{Lavender!40}12.1 &\cellcolor{Lavender!40}6.87 &\cellcolor{Lavender!40}4.09 &\cellcolor{Lavender!40}2.55 &\cellcolor{Lavender!40}1.62 &\cellcolor{Lavender!40}1.07 &\cellcolor{Lavender!40}0.70 &\cellcolor{Lavender!40}0.48 \\
& Scaled Addition & 15.1 & 8.54 & 5.32 & 3.18 & 1.97 & 1.32 & 0.82 & 0.26 \\
%& Approx. Add & 21.3 & 13.8 & 10.3 & 8.22 & 7.22 & 6.86 & 6.48 & 6.41 \\
&\cellcolor{Lavender!40}Abs. Subtraction &\cellcolor{Lavender!40}10.3 &\cellcolor{Lavender!40}5.35 &\cellcolor{Lavender!40}2.88 &\cellcolor{Lavender!40}1.43 &\cellcolor{Lavender!40}0.68 &\cellcolor{Lavender!40}0.42 &\cellcolor{Lavender!40}0.20 &\cellcolor{Lavender!40}0.10 \\
& Division & 18.3 & 13.6 & 10.54 & 9.50 & 8.74 & 8.82 & 8.31 & 8.39 \\
&\cellcolor{Lavender!40}Minimum &\cellcolor{Lavender!40}13.3 &\cellcolor{Lavender!40}6.92 &\cellcolor{Lavender!40}3.81 &\cellcolor{Lavender!40}1.96 &\cellcolor{Lavender!40}0.93 &\cellcolor{Lavender!40}0.57 &\cellcolor{Lavender!40}0.23 &\cellcolor{Lavender!40}0.15 \\
& Maximum & 9.42 & 4.96 & 2.64 & 1.38 & 0.68 & 0.37 & 0.14 & 0.11 \\
\midrule

% Data Section 9
\multirow{7}{*}{\begin{tabular}[c]{@{}l@{}}QRNG\\(Sobol)\end{tabular}}
&\cellcolor{Lavender!40}Multiplication &\cellcolor{Lavender!40}9.11 &\cellcolor{Lavender!40}3.88 &\cellcolor{Lavender!40}1.98 &\cellcolor{Lavender!40}1.05 &\cellcolor{Lavender!40}0.58 &\cellcolor{Lavender!40}0.27 &\cellcolor{Lavender!40}0.14 &\cellcolor{Lavender!40}0.07 \\
& Scaled Addition & 7.55 & 3.79 & 2.59 & 0.94 & 0.47 & 0.33 & 0.12 & 0.06 \\
%& Approx. Add & 4.16 & 2.08 & 1.04 & 0.52 & 0.26 & 0.13 & 0.06 & 0.03 \\
&\cellcolor{Lavender!40}Abs. Subtraction&\cellcolor{Lavender!40}4.16 &\cellcolor{Lavender!40}2.08 &\cellcolor{Lavender!40}1.04 &\cellcolor{Lavender!40}0.52 &\cellcolor{Lavender!40}0.26 &\cellcolor{Lavender!40}0.13 &\cellcolor{Lavender!40}0.06 &\cellcolor{Lavender!40}0.03 \\
& Division & 8.34 & 4.86 & 3.26 & 2.86 & 2.71 & 2.72 & 2.76 & 2.75 \\
&\cellcolor{Lavender!40}Minimum &\cellcolor{Lavender!40}6.51 &\cellcolor{Lavender!40}3.19 &\cellcolor{Lavender!40}1.57 &\cellcolor{Lavender!40}0.79 &\cellcolor{Lavender!40}0.39 &\cellcolor{Lavender!40}0.20 &\cellcolor{Lavender!40}0.10 &\cellcolor{Lavender!40}0.05 \\
& Maximum & 5.99 & 3.09 & 1.56 & 0.77 & 0.39 & 0.20 & 0.10 & 0.05 \\
\bottomrule
\end{tabular}
}
\label{tab:sc_arithmetic_comparison}
\vspace{-1em}
\end{table}

\subsection{SC Arithmetic Operations}
%{\textbf{\textit{4.2. SC Arithmetic Operations:}}} 
We next measure the MAE of common SC operations: multiplication (using \texttt{AND}), scaled addition (using \texttt{MUX}), absolute subtraction (using \texttt{XOR}), division (using the design proposed in~\cite{CORDIV}), and minimum and maximum~\cite{Sorting_Alam_2022}. Table~\ref{tab:sc_arithmetic_comparison} reports the results. As can be seen, the three proposed %TRNG-based 
methods consistently produce accuracy results that are close to the software-based MATLAB generator, confirming their effectiveness for true-random bit-stream generation. The block-extraction method achieves the lowest MAEs among the proposed designs, closely tracking MATLAB for multiplication, scaled addition, and minimum/maximum operations as the bit-stream length increases. The counter-logic method also delivers strong accuracy, with results nearly indistinguishable from the block extraction method in many cases. The majority-logic converter starts with higher error at short lengths but converges to MATLAB-level accuracy as 
$N$ grows. 
These results demonstrate that the proposed designs %TRNG-based generators
maintain reliable accuracy for all essential SC arithmetic operations. The consistent convergence of MAE with increasing bit-stream length confirms that the statistical quality of the proposed converters is sufficient to support a wide range of SC arithmetic.

\begin{table}[!t]
\centering
\caption{STD of SCC Values for \underline{Independent} Bit-stream Generation (Average SCC=0)}
\vspace{-0.75em}
\setlength{\tabcolsep}{2pt}
\renewcommand{\arraystretch}{0.95}
% Use the D column type for decimal alignment
%\begin{tabular}{l l *{8}{D{.}{.}{2.3}}}
\resizebox{\columnwidth}{!}{
\begin{tabular}{l c *{8}{c}}
\toprule
\multirow{2}{*}{\textbf{SBS Generation Method}} & \multirow{2}{*}{\begin{tabular}[c]{@{}c@{}}\textbf{Block}\\\textbf{size}\textbf{(M)} \end{tabular}} & \multicolumn{8}{c}{\textbf{Bit-stream length (N)}} \\
\cmidrule(lr){3-10}
% Wrap each header number in \multicolumn to center it
& & \multicolumn{1}{c}{\textbf{8}} & \multicolumn{1}{c}{\textbf{16}} & \multicolumn{1}{c}{\textbf{32}} & \multicolumn{1}{c}{\textbf{64}} & \multicolumn{1}{c}{\textbf{128}} & \multicolumn{1}{c}{\textbf{256}} & \multicolumn{1}{c}{\textbf{512}} & \multicolumn{1}{c}{\textbf{1024}} \\
\midrule

\multirow{8}{*}{\begin{tabular}[c]{@{}c@{}} %\textbf{Proposed Method \ding{172}} \\ 
\textbf{\textit{\ding{172}~Block Extraction}}\\+\textbf{\textcolor{mysecpurple}{TRNG1}}~\cite{TRNG_randomorg} \end{tabular}} &\cellcolor{Lavender!40}3 &\cellcolor{Lavender!40}0.62 &\cellcolor{Lavender!40}0.57 &\cellcolor{Lavender!40}0.45 &\cellcolor{Lavender!40}0.31 &\cellcolor{Lavender!40}0.21 &\cellcolor{Lavender!40}0.15 &\cellcolor{Lavender!40}0.10 &\cellcolor{Lavender!40}0.07 \\
%\rowcolor[HTML]{EFEFEF}
& 4 & 0.65 & 0.61 & 0.52 & 0.42 & 0.29 & 0.21 & 0.15 & 0.10 \\
&\cellcolor{Lavender!40}5 &\cellcolor{Lavender!40}0.64 &\cellcolor{Lavender!40}0.62 &\cellcolor{Lavender!40}0.55 &\cellcolor{Lavender!40}0.45 &\cellcolor{Lavender!40}0.34 &\cellcolor{Lavender!40}0.26 &\cellcolor{Lavender!40}0.18 &\cellcolor{Lavender!40}0.13 \\
%\rowcolor[HTML]{EFEFEF}
& 6 & 0.64 & 0.62 & 0.56 & 0.47 & 0.37 & 0.29 & 0.22 & 0.15 \\
&\cellcolor{Lavender!40}7 &\cellcolor{Lavender!40}0.65 &\cellcolor{Lavender!40}0.62 &\cellcolor{Lavender!40}0.55 &\cellcolor{Lavender!40}0.48 &\cellcolor{Lavender!40}0.39 &\cellcolor{Lavender!40}0.31 &\cellcolor{Lavender!40}0.23 &\cellcolor{Lavender!40}0.18 \\
%\rowcolor[HTML]{EFEFEF}
& 8 & 0.65 & 0.62 & 0.56 & 0.47 & 0.39 & 0.32 & 0.25 & 0.18 \\
&\cellcolor{Lavender!40}9 &\cellcolor{Lavender!40}0.65 &\cellcolor{Lavender!40}0.63 &\cellcolor{Lavender!40}0.56 &\cellcolor{Lavender!40}0.47 &\cellcolor{Lavender!40}0.39 &\cellcolor{Lavender!40}0.31 &\cellcolor{Lavender!40}0.25 &\cellcolor{Lavender!40}0.19 \\
%\rowcolor[HTML]{EFEFEF}
& 10 & 0.65 & 0.62 & 0.55 & 0.48 & 0.40 & 0.32 & 0.25 & 0.20 \\
%\addlinespace % Adds a little extra vertical space for separation
\midrule
\multirow{8}{*}{\begin{tabular}[c]{@{}c@{}} %\textbf{Proposed Method \ding{172}} \\
\textbf{\textit{\ding{172}~Block Extraction}}\\+\textbf{\textcolor{mysecpurple}{TRNG2}}~\cite{TRNG_randombeacons} \end{tabular}} &\cellcolor{Lavender!40}3 &\cellcolor{Lavender!40}0.62 &\cellcolor{Lavender!40}0.57 &\cellcolor{Lavender!40}0.45 &\cellcolor{Lavender!40}0.31 &\cellcolor{Lavender!40}0.21 &\cellcolor{Lavender!40}0.15 &\cellcolor{Lavender!40}0.11 &\cellcolor{Lavender!40}0.07 \\
& 4 & 0.64 & 0.61 & 0.52 & 0.42 & 0.29 & 0.21 & 0.15 & 0.10 \\
&\cellcolor{Lavender!40}5 &\cellcolor{Lavender!40}0.64 &\cellcolor{Lavender!40}0.63 &\cellcolor{Lavender!40}0.55 &\cellcolor{Lavender!40}0.46 &\cellcolor{Lavender!40}0.35 &\cellcolor{Lavender!40}0.25 &\cellcolor{Lavender!40}0.19 &\cellcolor{Lavender!40}0.13 \\
& 6 & 0.65 & 0.62 & 0.56 & 0.47 & 0.37 & 0.29 & 0.22 & 0.15 \\
&\cellcolor{Lavender!40}7 &\cellcolor{Lavender!40}0.65 &\cellcolor{Lavender!40}0.63 &\cellcolor{Lavender!40}0.56 &\cellcolor{Lavender!40}0.47 &\cellcolor{Lavender!40}0.39 &\cellcolor{Lavender!40}0.31 &\cellcolor{Lavender!40}0.24 &\cellcolor{Lavender!40}0.18 \\
& 8 & 0.66 & 0.63 & 0.56 & 0.47 & 0.39 & 0.31 & 0.24 & 0.18 \\
&\cellcolor{Lavender!40}9 &\cellcolor{Lavender!40}0.65 &\cellcolor{Lavender!40}0.62 &\cellcolor{Lavender!40}0.56 &\cellcolor{Lavender!40}0.47 &\cellcolor{Lavender!40}0.39 &\cellcolor{Lavender!40}0.31 &\cellcolor{Lavender!40}0.25 &\cellcolor{Lavender!40}0.19 \\
& 10 & 0.65 & 0.63 & 0.56 & 0.47 & 0.40 & 0.31 & 0.25 & 0.20 \\
\midrule

% Other data rows
\begin{tabular}[c]{@{}c@{}}  %\textbf{Proposed Method \ding{174}}\\ 
\textbf{\textit{\ding{173}~Counter Logic}+\textcolor{mysecpurple}{TRNG1}}~\cite{TRNG_randomorg} \end{tabular} & \cellcolor{Lavender!40}{--} &\cellcolor{Lavender!40}0.64 &\cellcolor{Lavender!40}0.57 &\cellcolor{Lavender!40}0.48 &\cellcolor{Lavender!40}0.41 &\cellcolor{Lavender!40}0.36 &\cellcolor{Lavender!40}0.31 &\cellcolor{Lavender!40}0.29 &\cellcolor{Lavender!40}0.27 \\
%\midrule
\textbf{\textit{\ding{173}~Counter Logic}+\textcolor{mysecpurple}{TRNG2}}~\cite{TRNG_randombeacons} & {--} & 0.63 & 0.57 & 0.48 & 0.41 & 0.35 & 0.32 & 0.29 & 0.28 \\
%\addlinespace
\midrule
\textbf{\textit{\ding{174}~Majority Logic}+\textcolor{mysecpurple}{TRNG1}}~\cite{TRNG_randomorg} & \cellcolor{Lavender!40}{--} &\cellcolor{Lavender!40}0.62 &\cellcolor{Lavender!40}0.61 &\cellcolor{Lavender!40}0.55 &\cellcolor{Lavender!40}0.47 &\cellcolor{Lavender!40}0.39 &\cellcolor{Lavender!40}0.31 &\cellcolor{Lavender!40}0.25 &\cellcolor{Lavender!40}0.20 \\
\textbf{\textit{\ding{174}~Majority Logic}+\textcolor{mysecpurple}{TRNG2}}~\cite{TRNG_randombeacons} & {--} & 0.62 & 0.61 & 0.55 & 0.47 & 0.38 & 0.32 & 0.25 & 0.20 \\
%\midrule

% Final data rows
%\texttt{MATLAB} (rand) & {--} & 11.2 & 7.84 & 5.54 & 3.91 & 2.76 & 1.96 & 1.38 & 0.97 \\
%PRNG ($\log_2{N}$-bit LFSR) & {--} & 11.8 & 6.10 & 3.17 & 1.51 & 0.77 & 0.39 & 0.21 & 0.10 \\
%QRNG (Sobol) & {--} & 6.25 & 3.13 & 1.56 & 0.78 & 0.39 & 0.19 & 0.09 & 0.04 \\
\bottomrule
%\multicolumn{10}{l}{\scriptsize For PRNG, we use a maximal length LFSR with polynomial $x^8+x^5+x^3+1$.} % Table note, left-aligned
\end{tabular}
}
\label{tab:scc_uncorrelated_bs}
\vspace{-0.75em}
\end{table}

\subsection{Bit-stream Correlation}
%{\textbf{\textit{4.3. SBS Correlation:}}}
We then evaluate the performance of the proposed methods in generating \textit{independent} and \textit{correlated}  bit-streams. The correlation between two bit-streams $X$ and $Y$ is quantified
using the SC Correlation (SCC) metric~\cite{Alaghi_SCC}:
\vspace{-0.25em}
$$
SCC(X,Y) = \left\{
	\begin{array}{llr}
		\frac{ad-bc}{N \times min(a+b,a+c)-(a+b)(a+c)} & ad>bc \\
		\frac{ad-bc}{(a+b)(a+c)-N\times max(a-d,0)} & else  
	\end{array} \right. \label{equ:S_{n,k}}
$$
\noindent where, $a$ counts positions where both bit-streams ($X$ and $Y$) are 1, $b$ where $X$=1 and $Y$=0, $c$ where $X$=0 and $Y$=1, and $d$ where both are 0~\cite{AygunAgile}. An SCC of 0 indicates the streams are uncorrelated and an SCC of 1 means they are maximally correlated.
%To measure the amount of independence (i.e., uncorrelation) among bit-streams, we utilize two distinct subsets of TRNGs for each proposed approach.
Our evaluations showed SCC values near 0 for all three methods when generating independent bit-streams. % as described in Section~\ref{prop}. 
Table~\ref{tab:scc_uncorrelated_bs} reports the standard deviation (STD) of the SCC values %between two pairs of bit-streams in %for a Monte Carlo simulation of 
over 100,000 iterations. As seen in the results, in all cases, the STD decreases as the bit-stream length increases, indicating that the SCC values become more tightly concentrated around 0. 
%To obtain uncorrelated pairs of bit-streams, we select completely non-overlapped sub-streams from the original TRNG stream. As the bit-stream length increases, the SCC values moves towards 0. The zero value of SCC among bit-stream pairs justifies the inherent independence among any sub-stream of TRNG source with an equidistribution of 0's and 1's.
%Another tightly coupled metric related to the degree of dependence between two random variables is Mutual Information (MI) which is widely used in information theory~\cite{Mutual_Info_PNAS2014}
Table~\ref{tab:scc_correlated_bs} further reports the mean and STD of the SCC values when generating \textit{correlated} bit-stream pairs using the proposed methods, as described in Section~\ref{prop}. As can be seen, in all cases, increasing the bit-stream length drives the SCC values toward the ideal value of 1.0 (indicating strong correlation) while simultaneously reducing the STD. For the \textit{Block Extraction} method, SCC values near 1.0 are achieved with block sizes larger than 5 for $N \geq 64$. 
\color{black}

\begin{table}[!t]
\centering
\caption{\textit{Avg/STD} of SCC Values for \underline{Correlated} Bit-stream Generation}
\vspace{-0.75em}
\setlength{\tabcolsep}{2.1pt}
\renewcommand{\arraystretch}{1.}
% Use the D column type for decimal alignment
%\begin{tabular}{l l *{8}{D{.}{.}{2.3}}}
\resizebox{\columnwidth}{!}{
\begin{tabular}{l c *{8}{c}}
\toprule
\multirow{2}{*}{\begin{tabular}[c]{@{}l@{}}\textbf{SBS Generation}\\ \textbf{Method} \end{tabular}} & \multirow{2}{*}{\begin{tabular}[c]{@{}c@{}}\textbf{Block}\\\textbf{size}\textbf{(M)} \end{tabular}} & \multicolumn{8}{c}{\textbf{Bit-stream length (N)}} \\
\cmidrule(lr){3-10}
% Wrap each header number in \multicolumn to center it
& & \multicolumn{1}{c}{\textbf{8}} & \multicolumn{1}{c}{\textbf{16}} & \multicolumn{1}{c}{\textbf{32}} & \multicolumn{1}{c}{\textbf{64}} & \multicolumn{1}{c}{\textbf{128}} & \multicolumn{1}{c}{\textbf{256}} & \multicolumn{1}{c}{\textbf{512}} & \multicolumn{1}{c}{\textbf{1024}} \\
\midrule

\multirow{8}{*}{\begin{tabular}[c]{@{}c@{}} %\textbf{Proposed Method \ding{172}} \\ 
\textbf{\textit{\ding{172}~Block}}\\\textbf{\textit{Extraction}}\\+\textbf{\textcolor{mysecpurple}{TRNG1}}~\cite{TRNG_randomorg} \end{tabular}} &\cellcolor{Lavender!40}3 & \cellcolor{Lavender!40}0.60/0.49 & \cellcolor{Lavender!40}0.72/0.45 & \cellcolor{Lavender!40}0.76/0.43 &\cellcolor{Lavender!40}0.76/0.42 &\cellcolor{Lavender!40}0.77/0.42 &\cellcolor{Lavender!40}0.76/0.42 &\cellcolor{Lavender!40}0.76/0.43 &\cellcolor{Lavender!40}0.77/0.42 \\
%\rowcolor[HTML]{EFEFEF}
& 4 & 0.62/0.49 & 0.76/0.43 & 0.84/0.36 & 0.87/0.33 & 0.87/0.33 & 0.87/0.33 & 0.88/0.33 & 0.88/0.33\\
&\cellcolor{Lavender!40}5 &\cellcolor{Lavender!40}0.62/0.49 &\cellcolor{Lavender!40}0.78/0.41 &\cellcolor{Lavender!40}0.88/0.33 &\cellcolor{Lavender!40}0.92/0.26 &\cellcolor{Lavender!40}0.94/0.24 &\cellcolor{Lavender!40}0.94/0.23 &\cellcolor{Lavender!40}0.94/0.24 &\cellcolor{Lavender!40}0.94/0.24 \\
%\rowcolor[HTML]{EFEFEF}
& 6 & 0.62/0.49 & 0.78/0.41 & 0.88/0.32 & 0.94/0.24 & 0.96/0.19 & 0.97/0.17 & 0.97/0.17 & 0.97/0.17 \\
&\cellcolor{Lavender!40}7 &\cellcolor{Lavender!40}0.62/0.48 &\cellcolor{Lavender!40}0.79/0.41 &\cellcolor{Lavender!40}0.89/0.31 &\cellcolor{Lavender!40}0.94/0.23 &\cellcolor{Lavender!40}0.97/0.18 &\cellcolor{Lavender!40}0.98/0.14 &\cellcolor{Lavender!40}0.98/0.13 &\cellcolor{Lavender!40}0.98/0.12 \\
%\rowcolor[HTML]{EFEFEF}
& 8 & 0.62/0.48 & 0.80/0.40 & 0.89/0.32 & 0.94/0.24 & 0.97/0.17 & 0.98/0.13 & 0.99/0.11 & 0.99/0.08 \\
&\cellcolor{Lavender!40}9 &\cellcolor{Lavender!40}0.62/0.48 &\cellcolor{Lavender!40}0.78/0.41 &\cellcolor{Lavender!40}0.88/0.32 &\cellcolor{Lavender!40}0.94/0.23 &\cellcolor{Lavender!40}0.97/0.17 &\cellcolor{Lavender!40}0.98/0.13 &\cellcolor{Lavender!40}0.99/0.09 &\cellcolor{Lavender!40}0.99/0.07 \\
%\rowcolor[HTML]{EFEFEF}
& 10 & 0.63/0.48 & 0.78/0.41 & 0.88/0.33 & 0.94/0.23 & 0.97/0.17 & 0.98/0.12 & 0.99/0.07 & 1.00/0.06 \\
%\addlinespace % Adds a little extra vertical space for separation
\midrule
\multirow{8}{*}{\begin{tabular}[c]{@{}c@{}} %\textbf{Proposed Method \ding{172}} \\
\textbf{\textit{\ding{172}~Block}}\\\textbf{\textit{Extraction}}\\+\textbf{\textcolor{mysecpurple}{TRNG2}}~\cite{TRNG_randombeacons} \end{tabular}} &\cellcolor{Lavender!40}3 &\cellcolor{Lavender!40}0.59/0.49 &\cellcolor{Lavender!40}0.71/0.45 &\cellcolor{Lavender!40}0.75/0.43 &\cellcolor{Lavender!40}0.76/0.42 &\cellcolor{Lavender!40}0.77/0.42 &\cellcolor{Lavender!40}0.77/0.42 &\cellcolor{Lavender!40}0.77/0.42 &\cellcolor{Lavender!40}0.76/0.42 \\
& 4 & 0.61/0.49 & 0.76/0.43 & 0.85/0.36 & 0.87/0.33 & 0.87/0.33 & 0.88/0.32 & 0.88/0.32 & 0.88/0.33 \\
&\cellcolor{Lavender!40}5 &\cellcolor{Lavender!40}0.62/0.49 &\cellcolor{Lavender!40}0.78/0.41 &\cellcolor{Lavender!40}0.88/0.33 &\cellcolor{Lavender!40}0.92/0.27 &\cellcolor{Lavender!40}0.94/0.24 &\cellcolor{Lavender!40}0.94/0.23 &\cellcolor{Lavender!40}0.94/0.23 &\cellcolor{Lavender!40}0.94/0.24 \\
& 6 & 0.62/0.48 & 0.79/0.41 & 0.88/0.32 & 0.93/0.25 & 0.96/0.19 & 0.97/0.18 & 0.97/0.18 & 0.97/0.18 \\
&\cellcolor{Lavender!40}7 &\cellcolor{Lavender!40}0.62/0.48 &\cellcolor{Lavender!40}0.78/0.42 &\cellcolor{Lavender!40}0.88/0.32 &\cellcolor{Lavender!40}0.94/0.24 &\cellcolor{Lavender!40}0.97/0.18 &\cellcolor{Lavender!40}0.98/0.13 &\cellcolor{Lavender!40}0.98/0.12 &\cellcolor{Lavender!40}0.98/0.12 \\
& 8 & 0.62/0.48 & 0.77/0.42 & 0.89/0.32 & 0.94/0.23 & 0.97/0.17 & 0.98/0.13 & 0.99/0.10 & 0.99/0.09 \\
&\cellcolor{Lavender!40}9 &\cellcolor{Lavender!40}0.62/0.48 &\cellcolor{Lavender!40}0.78/0.41 &\cellcolor{Lavender!40}0.88/0.32 &\cellcolor{Lavender!40}0.94/0.24 &\cellcolor{Lavender!40}0.97/0.17 &\cellcolor{Lavender!40}0.98/0.12 &\cellcolor{Lavender!40}0.99/0.09 &\cellcolor{Lavender!40}0.99/0.07 \\
& 10 & 0.62/0.49 & 0.79/0.41 & 0.88/0.32 & 0.94/0.24 & 0.97/0.17 & 0.99/0.12 & 0.99/0.10 & 1.00/0.06 \\
\midrule

% Other data rows
\begin{tabular}[c]{@{}c@{}}  %\textbf{Proposed Method \ding{174}}\\ 
\textbf{\textit{\ding{173}~Counter Logic}}\\ +\textbf{\textcolor{mysecpurple}{TRNG1}}~\cite{TRNG_randomorg} \end{tabular} & \cellcolor{Lavender!40}{--} &\cellcolor{Lavender!40}0.62/0.48 &\cellcolor{Lavender!40}0.75/0.43 &\cellcolor{Lavender!40}0.83/0.38 &\cellcolor{Lavender!40}0.89/0.31 &\cellcolor{Lavender!40}0.92/0.26 &\cellcolor{Lavender!40}0.95/0.22 &\cellcolor{Lavender!40}0.96/0.19 &\cellcolor{Lavender!40}0.97/0.16 \\
%\midrule
\begin{tabular}[c]{@{}c@{}} \textbf{\textit{\ding{173}~Counter Logic}}\\ +\textbf{\textcolor{mysecpurple}{TRNG2}}~\cite{TRNG_randombeacons} \end{tabular} & {--} & 0.62/0.48 & 0.76/0.43 & 0.83/0.37 & 0.88/0.32 & 0.92/0.27 & 0.94/0.23 & 0.96/0.19 & 0.97/0.16 \\
%\addlinespace
\midrule
\begin{tabular}[c]{@{}c@{}} \textbf{\textit{\ding{174}~Majority Logic}}\\ +\textbf{\textcolor{mysecpurple}{TRNG1}}~\cite{TRNG_randomorg}\end{tabular} & \cellcolor{Lavender!40}{--} &\cellcolor{Lavender!40}0.58/0.49 &\cellcolor{Lavender!40}0.77/0.42 &\cellcolor{Lavender!40}0.87/0.33 &\cellcolor{Lavender!40}0.93/0.25 &\cellcolor{Lavender!40}0.96/0.18 &\cellcolor{Lavender!40}0.98/0.13 &\cellcolor{Lavender!40}0.99/0.08 &\cellcolor{Lavender!40}1.00/0.06 \\
\begin{tabular}[c]{@{}c@{}} \textbf{\textit{\ding{174}~Majority Logic}}\\ +\textbf{\textcolor{mysecpurple}{TRNG2}}~\cite{TRNG_randombeacons} \end{tabular} & {--} & 0.59/0.49 & 0.76/0.43 & 0.87/0.33 & 0.93/0.25 & 0.97/0.18 & 0.98/0.13 & 0.99/0.09 & 1.00/0.06 \\
%\midrule

% Final data rows
%\texttt{MATLAB} (rand) & {--} & 11.2 & 7.84 & 5.54 & 3.91 & 2.76 & 1.96 & 1.38 & 0.97 \\
%PRNG ($\log_2{N}$-bit LFSR) & {--} & 11.8 & 6.10 & 3.17 & 1.51 & 0.77 & 0.39 & 0.21 & 0.10 \\
%QRNG (Sobol) & {--} & 6.25 & 3.13 & 1.56 & 0.78 & 0.39 & 0.19 & 0.09 & 0.04 \\
\bottomrule
%\multicolumn{10}{l}{\scriptsize For PRNG, we use a maximal length LFSR with polynomial $x^8+x^5+x^3+1$.} % Table note, left-aligned
\end{tabular}
}
\label{tab:scc_correlated_bs}
%\vspace{-1.5em}
\end{table}

\section{Hardware Cost Comparison}
\label{hard}
We next compare the hardware costs of the three proposed converters with the SOTA LFSR- and Sobol-based generators. 
The comparison separates the cost of converting a true-random stream from the cost of generating a pseudo- or quasi-random one.
The proposed converters need a true-random source, but that source is not part of the converter: it can be built in any of several device technologies, many of them non-CMOS, and it can be far cheaper than a PRNG or QRNG.  
%To ensure a fair and direct comparison, we distinguish between the conversion cost of our proposed designs and the generation cost of conventional PRNG/QRNG-based approaches. While our methods require a true-random source to supply unbiased input bits, such sources can be realized using a variety of device technologies (often non-CMOS) and are not intrinsic to the converter design itself. Compact, low-energy TRNGs have been proposed with orders of magnitude lower cost than PRNGs and QRNGs. 
The superparamagnetic tunnel junction TRNG of \cite{TRNG_2017}, for example, occupies only 2 $\mu m^2$ and consumes 20 $fJ$ %energy 
per bit.
%SFC-TRNG generates high-quality true-random bits with a compact footprint of 11.17 $\mu m^2$ and an energy cost of only 3.98 $pJ$ per bit.
Table~\ref{hw_cost_RNG} therefore reports the cost of the CMOS conversion circuits of Fig. ~\ref{fig:bs_trng} on their own, so the numbers hold for any TRNG that feeds them. 
%For portability and fair comparison, Table~\ref{hw_cost_RNG} reports the cost of the CMOS conversion circuits shown in Figure~\ref{fig:bs_trng}, independent of any %This approach highlights the efficiency of our proposed architectures, independent of the 
%specific TRNG implementation.
%of the total cost of SBS generation between the proposed converters and the SOTA PRNG/QRNG methods, we need to consider the cost of TRNG generation in our methods. However, our proposed converters are compatible with any true-random source of generating bits, some of which are non-CMOS in other technologies. So, for the proposed methods, we only report the cost of conversion circuits shown in Fig.~\ref{fig:bs_trng}. 
We described all designs in Verilog and 
%The hardware performance of the proposed SBS generation methods was evaluated through Verilog HDL implementation. We 
synthesized them with the Synopsys Design Compiler v2024.09-SP4 using the 45$nm$ FreePDK gate library~\cite{FreePDK45}.
%The synthesis results were obtained using the Synopsys Design Compiler (W-2024.09-SP4) targeting a 45-nm CMOS technology. 
For the PRNG generator, we implemented an 8-bit maximal-period LFSR with feedback polynomial $x^{8}+x^{6}+x^{3} +x^ {2} +1$ followed by an 8-bit comparator. 
%The LFSR implementation utilizes the maximal-length polynomial $x^{8}+x^{6}+x^{3}+x^2+1$. 
For the QRNG generator, we implemented two representative SOTA designs: I) a comparator-based generator built from an 8-bit Sobol sequence generator with an 8-bit comparator~\cite{liu2017energy}, and II) an FSM-based generator with 8-bit precision as described in~\cite{Najafi_FSM_DATE2021}.
The three converters are synthesized at the same 8-bit data precision, enabling a consistent comparison across all designs.

Table~\ref{hw_cost_RNG} summarizes the hardware cost results. %The results highlight that 
The \textit{block-extraction} and \textit{majority-logic} converters are the two smallest designs at 72~$\mu$m$^2$ and 61~$\mu$m$^2$, with critical paths near 0.4~ns and energy of 0.02~pJ and 0.03~pJ per cycle. %methods achieve very compact designs with low area, short critical path latency, and minimal energy per cycle, making them lightweight options for integration into SC systems. 
Both undercut the LFSR-based generator by more than 3$\times$ in area and more than 10$\times$ in energy per cycle, and undercut both Sobol-based generators by wider margins. 
The \textit{counter-logic} converter  is larger and slower than the LFSR baseline because of its $n$-bit counter, \texttt{MUX}, and loading logic, though it uses fewer bits from the source than the other two converters.
%while functionally effective, incurs higher area and power overheads due to its more complex circuitry. 
The three designs thus span a range of trade-offs: \textit{block extraction} and \textit{majority logic} for area and energy, \textit{counter logic} where source bandwidth is the constraint.
%Overall, the data confirm that the proposed converters can be realized with modest CMOS resources, and the diversity of trade-offs across the three designs enables flexibility based on application needs. 
%When compared to LFSR- and Sobol-based generators, the \textit{block-extraction} and \textit{majority-logic} designs report lower area, area–delay products, and energy per cycle, suggesting favorable efficiency. %However, we emphasize that for an accurate one-to-one comparison, the cost of the TRNG itself must be included alongside the cost of our conversion circuits.
%A recent TRNG design synthesized in 45nm technology was reported to occupy an area of $223.39\mu m^2$ and consume $34\mu W$ power and $34fJ$ energy per random bit.

%\blue{Table~\ref{hw_cost_RNG} reports the hardware cost numbers. The block extraction and majority logic pipeline designs provides far less energy consumption compared to the conventional SOTA designs. However, the counter logic method consumes vast majority of hardware resources due to its more complex structure compared to the other designs.
%As summarized in Table~\ref{hw_cost_RNG}, the proposed architectures are benchmarked against two conventional generators: a Finite-State Machine (FSM) based Sobol sequence generator~\cite{Najafi_FSM_DATE2021} and a LFSR. 
%}

\begin{table}[!t]
%\color{violet}
%\vspace{-1em}
\centering
\caption{Hardware implementation cost of SBS generation methods. The designs are synthesized in 45$nm$ for 8-bit precision.}
%\begin{threeparttable}
\vspace{-0.75em}
\setlength{\tabcolsep}{5pt}
\renewcommand{\arraystretch}{0.9}
\resizebox{\columnwidth}{!}{
\begin{tabular}{lccccc} 
\toprule
%\textbf{SBS Generation Method} 
\begin{tabular}[l]{@{}l@{}}\textbf{SBS Generation} \\\textbf{Method}\end{tabular}
& \begin{tabular}[c]{@{}c@{}}\textbf{Area} \\($\mu m^2$)\end{tabular} & \begin{tabular}[c]{@{}c@{}}\textbf{Power@}\\\textbf{Max Freq.} \\($\mu W$)\end{tabular} & \begin{tabular}[c]{@{}c@{}}\textbf{CPL} \\($ns$)\end{tabular} & \begin{tabular}[c]{@{}c@{}}\textbf{Energy}\\ \textbf{per Cycle} \\($pJ$)\end{tabular} & \begin{tabular}[c]{@{}c@{}}\textbf{Area$\times$Delay} \\($\mu m^2\times ns$)\end{tabular} \\ 
\midrule
\textbf{\ding{172} \textit{Block Extraction}}& 72 & \underline{\textbf{57.1}} & 0.41 & \underline{\textbf{0.02}} & 29.5\\
%\textbf{\ding{172} \textit{Block Extraction}}--\textbf{2\texttimes parallel} & 136 & 107.4 & 0.37 & 0.04 & 50.3\\
%\textbf{\ding{172} \textit{Block Extraction}}--\textbf{4\texttimes parallel} & 260 & 209.4 & 0.38 & 0.08 & 98.8\\
\textbf{\ding{173} \textit{Counter Logic}} & 382 & 1568 & 0.39 & 0.61 & 148\\
\textbf{\ding{ 174} \textit{Majority Logic}} & \underline{\textbf{61}} & 85.7 & 0.42 & 0.03 & \underline{\textbf{25.6}} \\
\midrule
LFSR + Comparator & 233 & 949 & 0.37 & 0.35 & 86.2\\ 
Sobol + Comparator~\cite{liu2017energy} & 850 & 1307 & 0.45 & 0.59 & 382\\ 
%\cline{1-2}\cline{4-4}
Sobol (FSM-based)~\cite{Najafi_FSM_DATE2021} & 281 & 720 & 0.36 & 0.26 & 101\\ 
%\hline
\bottomrule
\end{tabular}
}
%\vspace{-1em}
%\parbox{\columnwidth}
\vspace{-0.75em}
\justify{\scriptsize{ 
The reported costs exclude the TRNG source, which is external to the converter design and can be implemented separately using low-cost non-CMOS technologies such as~\cite{TRNG_2017}.
%The cost of TRNG sources are not tied to the converter designs and are often realized by low-cost non-CMOS technologies such as~\cite{TRNG_2017}.
}}
\label{hw_cost_RNG}
%\vspace{-1em}
%\justify{\scriptsize{The maximal period with the polynomial $x^{8}+x^{6}+x^{3}+x^2+1$ is used for LFSR.}}
\vspace{-1.5em}
\end{table}

\section{Soft-Error Tolerance Comparison}
\label{Sec:softerror}

%Fault tolerance is one of the attractive properties of SC designs. The processing elements of SC systems inherently tolerate high rates of soft errors (i.e., bit flips) as they process data in the stochastic domain. However, the bit-stream generators that convert binary data to stochastic bit-streams are vulnerable to bit flips as they operate in the binary domain. 
A key advantage of SC is its robustness to transient faults, i.e., soft errors. 
%Since values are represented and processed as probabilities over long bit-streams, SC circuits can sustain a high rate of bit flips without significantly affecting computational outcomes. A flipped bit in an $N$-bit stream changes the encoded value by only $1/N$. 
Values are represented and processed as probabilities over long bit-streams, so a flipped bit in an $N$-bit stream changes the encoded value by only $1/N$, and an SC datapath sustains a high rate of bit flips with little effect on its output.
%This error-masking property, however, does not extend equally to all components of an SC system. In particular, the front-end SBS generators remain sensitive to soft errors, as their operation relies on conventional binary logic.
The front-end generator does not share this protection.  It is conventional binary logic, and a flip in its state can shift the probability of every bit that follows.
We therefore compare the soft-error tolerance of the three converters with the two Sobol-based generators when converting 8-bit values into 256-bit streams. 
%Here, we compare the soft error tolerance of the proposed TRNG-to-SBS converters and the SOTA Sobol-based SBS generators when converting 8-bit data into 256-bit SBS. We inject soft errors at different rates by flipping bits in the input and output bits of different components of the converters. 
We inject errors by flipping the input and output bits of each component of a design at rates from 1\% to 30\% and measure the resulting MAE.

Table~\ref{tab:noise_injection_rng} 
reports the results. %impact of the injected soft errors on the accuracy of different SBS generation methods when $N$=256. 
%Among the proposed converters, the
\textit{Block extraction} is the most tolerant of the three converters. Its error rises gradually with the injection rate and stays below 20\% at 30\% noise. 
%demonstrates the highest resilience, with errors growing gradually and remaining moderate even under high noise injection (e.g., 19.1\% MAE at 30\% noise). 
%\textit{Majority-logic} %degrades faster but remains well-behaved across the range.
\textit{Majority logic} shows a higher MAE rate, with accuracy decreasing more quickly than \textit{block extraction} but still maintaining consistent behavior across the entire noise range. 
%In contrast, the \textit{counter-logic} method is more sensitive to injected faults, reflecting the vulnerability of its sequential counting structure to error accumulation. 
\textit{Counter logic} is the most sensitive, with 14.3\% MAE at 1\% noise: a flip in the loaded count or the counter state changes the number of bits flipped for the rest of the window, so a single fault affects many output bits.
The lower block of Table~\ref{tab:noise_injection_rng} separates the effect of noise from each method's noise-free error. 
Measured this way, \textit{block extraction} is the most tolerant design in the comparison at low noise: its MAE rises by 0.17\% at 1\% injection and 0.64\% at 2\%, against 0.91\% and 1.41\% for the comparator-based Sobol generator. The two match at 5\% and 10\%, and \textit{block extraction} stays within 2.5\% of the Sobol design up to 30\%. It beats the FSM-based Sobol generator at every rate. The stateless comparator of \textit{block extraction} has no error to accumulate, whereas a flip in the Sobol generator's state corrupts every subsequent number. The lower absolute MAE of the Sobol designs in the upper block therefore reflects their lower starting error, not greater tolerance to faults, and it comes with deterministic, reproducible sequences. \textit{Block extraction} matches or exceeds their fault tolerance while producing streams that no observer can predict.

\begin{table}[!t]
\centering
\caption{MAE (\%) of SBS generation methods under injected soft errors ($N$=256). The lower block reports the increase over the noise-free MAE of each method.}
\vspace{-0.5em}
\resizebox{\columnwidth}{!}{
\renewcommand{\arraystretch}{0.9}
\begin{tabular}{l *{7}{c}}
\toprule
\multirow{2}{*}{\begin{tabular}[l]{@{}l@{}}\textbf{SBS Generation}\\ \textbf{Method} \end{tabular}} & \multicolumn{7}{c}{\textbf{Injected Noise (\%)}} \\
\cmidrule(lr){2-8}
& \textbf{0} & \textbf{1} & \textbf{2} & \textbf{5} & \textbf{10} & \textbf{20} & \textbf{30} \\
\midrule
\multicolumn{8}{l}{\textit{Absolute MAE (\%)}} \\
\textbf{\ding{172} \textit{Block Extraction}} & 1.97 & 2.14 & 2.61 & 4.35 & 7.51 & 13.5 & 19.1 \\
\textbf{\ding{173} \textit{Counter Logic}}    & 2.43 & 14.3 & 17.2 & 19.7 & 20.7 & 21.3 & 21.4 \\
\textbf{\ding{174} \textit{Majority Logic}}   & 2.35 & 4.06 & 6.26 & 11.6 & 16.9 & 22.1 & 24.4 \\
Sobol + Comparator~\cite{liu2017energy}      & 0.19 & 1.1  & 1.6  & 3.0  & 5.3  & 10.2 & 15.1 \\
Sobol (FSM based)~\cite{Najafi_FSM_DATE2021} & 0.19 & 1.4  & 2.6  & 6.0  & 10.8 & 17.7 & 21.8 \\
\midrule
\multicolumn{8}{l}{\textit{Increase over noise-free MAE (\%)}} \\
\textbf{\ding{172} \textit{Block Extraction}} & -- & 0.17 & 0.64 & 2.38 & 5.54 & 11.5 & 17.1 \\
\textbf{\ding{173} \textit{Counter Logic}}    & -- & 11.9 & 14.8 & 17.3 & 18.3 & 18.9 & 19.0 \\
\textbf{\ding{174} \textit{Majority Logic}}   & -- & 1.71 & 3.91 & 9.25 & 14.6 & 19.8 & 22.1 \\
Sobol + Comparator~\cite{liu2017energy}      & -- & 0.91 & 1.41 & 2.81 & 5.11 & 10.0 & 14.9 \\
Sobol (FSM based)~\cite{Najafi_FSM_DATE2021} & -- & 1.21 & 2.41 & 5.81 & 10.6 & 17.5 & 21.6 \\
\bottomrule
\end{tabular}
}
\label{tab:noise_injection_rng}
\end{table}

%\vspace{0.25em}
\section{Case Study: Bilinear Image Interpolation}
\label{case_study}
\begin{figure}[!t]
    \centering
    %\vspace{-0.5em}
    \includegraphics[width=\linewidth]{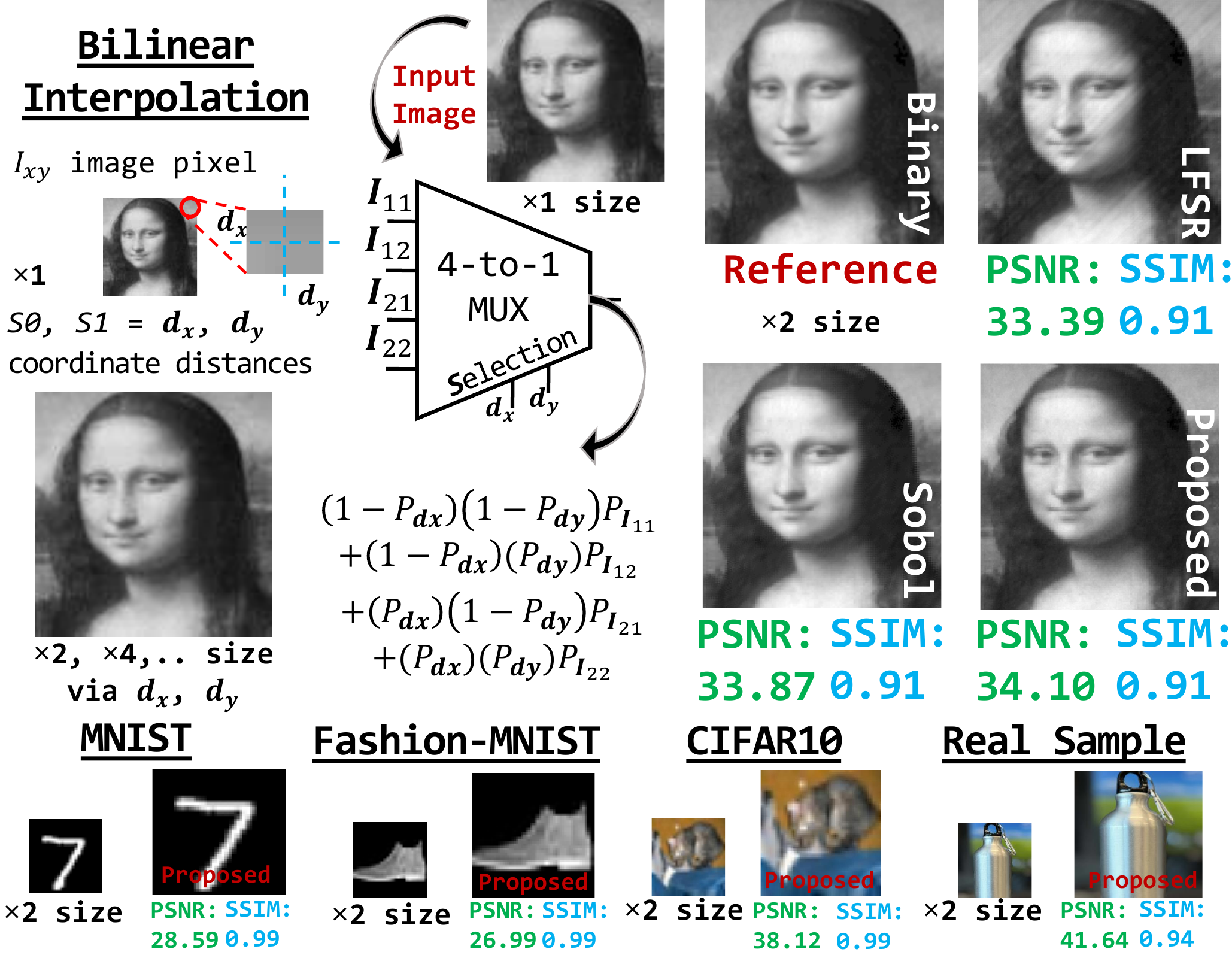}
    \vspace{-2em}
    \caption{SC circuit and output samples of image-processing bilinear interpolation ($N$ = 1024) using the proposed \textit{Block Extraction}. %Software-based case study on image interpolation. 
%The original study is inspired by prior work on stochastic sequence generators~\cite{moghadam2024p2lsg}.
}
    \label{fig:image_case_study}
    \vspace{-1.em}
\end{figure}

%{\textbf{\textit{4.4. SC Image Processing Case Study:}}} 
We further evaluate the performance using an %SC-based 
image processing case study. \textit{Bilinear interpolation} enlarges an image by estimating intermediate pixel values through two sequential linear interpolations along the \texttt{x}- and \texttt{y}-axes~\cite{9342625}. For any four neighboring pixels $I_{11}$, $I_{12}$, $I_{21}$, $I_{22}$, the value of a new pixel $I(x,y)$ is derived from their intensities and the relative offsets $(dx, dy)$ of the interpolation point: %The conventional relationship is: 
$I(x,y) = (1-dx)(1-dy)I_{11} + (1-dx)(dy)I_{12} + (dx)(1-dy)I_{21} + (dx)(dy)I_{22}$. In the stochastic domain, this operation can be realized with a 4-to-1 \texttt{MUX}~\cite{p2lsg_ASP_DAC24}, where the four pixel bit-streams act as inputs and the bit-stream representations of $dx$ and $dy$ serve as selection port signals. The resulting output bit-stream provides the estimated pixel intensity for the up-scaled image (e.g., from $\times1$ to $\times$2 in size). Fig.~\ref{fig:image_case_study} depicts the SC circuit and sample images. We compare the performance of three SBS generation methods: %To generate the SBS for each pixel, we apply three RNG approaches: 
LFSR (PRNG), Sobol (QRNG), and the proposed \textit{block extraction} method. The generated SBS are processed with the circuit %the 4-to-1 MUX 
to estimate the new pixel intensities for the enlarged image. Compared to the conventional binary baseline, the LFSR and Sobol methods achieved PSNRs of $33.39$ dB and $33.87$ dB, respectively, on the sample Mona Lisa image, with an SSIM of 0.91. %As expected, 
The proposed \textit{block extraction} method delivered a similar high accuracy, %the best performance,
achieving a PSNR of $34.10$ dB at the same SSIM level. %, representing a measurable improvement over both PRNG and QRNG approaches. 
%Achieving even fractional ($0.X$) PSNR gains is challenging in SC-based image processing, making this result particularly significant as it surpasses the well-known Sobol method.
Table~\ref{tab:sc_upsampling_comparison} extends the evaluation to %presents the results of the LFSR, Sobol, and proposed methods on 
the average of 10,000 test images from MNIST~\cite{lecun2002gradient}, Fashion-MNIST~\cite{xiao2017fashion}, and CIFAR-10~\cite{krizhevsky2009learning} using bit-streams with lengths of 512 and 1024. %The table includes two bit-stream lengths, 512 and 1024, so we can see how the amount of stochastic precision affects the performance. 
%As can be seen, 
Across all datasets, Sobol generally outperformed LFSR, showing the benefits of low-discrepancy sampling. %reduces randomness in the results. 
\textit{Block extraction} achieved the highest PSNR on MNIST and remained close to Sobol on Fashion-MNIST and CIFAR-10. SSIM scores were consistently high for all methods, between $0.98$ and $0.99$. %The table shows how each sampling strategy compares and how using a longer bit-stream improves the overall performance

These results indicate that true-random SBS from the proposed converters can replace deterministic PRNG or QRNG streams in an SC datapath with no change to the arithmetic and a small, bounded cost in accuracy.

%===================================
%   Mehran
%===================================
\begin{table}[!t]
\centering
%\vspace{-1em}
%\small
\caption{Performance of SC Bilinear Interpolation %on \textbf{10,000 test images} from MNIST, Fashion-MNIST, and CIFAR-10 
when generating SBS with LFSR, Sobol, and Proposed Block Extraction. %Results are reported for bit-stream lengths of $N=512$,$1024$. 
%PSNR in dB.
}
\vspace{-0.75em}
\label{tab:sc_upsampling_comparison}
\setlength{\tabcolsep}{2.5pt}
\resizebox{\columnwidth}{!}{
\begin{tabular}{l ccccccc}
\toprule
\multirow{3}{*}{\textbf{Dataset}} & \multirow{3}{*}{\begin{tabular}[c]{@{}c@{}} \textbf{Bit-stream}\\ \textbf{Length} \textbf{(N)} \end{tabular}} & \multicolumn{3}{c}{\textbf{PSNR (dB)}} & \multicolumn{3}{c}{\textbf{SSIM}} \\
\cmidrule(lr){3-5} \cmidrule(lr){6-8}
 & & \begin{tabular}[c]{@{}c@{}} \textbf{LFSR+}\\ \textbf{Comp.} \end{tabular} & \begin{tabular}[c]{@{}c@{}} \textbf{Sobol+}\\ \textbf{Comp.} \end{tabular} & \begin{tabular}[c]{@{}c@{}} \ding{172}\textbf{\textit{Block}}\\ \textbf{\textit{Extraction}} \end{tabular} & \begin{tabular}[c]{@{}c@{}} \textbf{LFSR+}\\ \textbf{Comp.} \end{tabular} & \begin{tabular}[c]{@{}c@{}} \textbf{Sobol+}\\ \textbf{Comp.} \end{tabular} & \begin{tabular}[c]{@{}c@{}} \ding{172}\textbf{\textit{Block}}\\ \textbf{\textit{Extraction}} \end{tabular} \\
\midrule
\multirow{2}{*}{\begin{tabular}[l]{@{}l@{}} \textbf{MNIST}\\ \cite{lecun2002gradient} \end{tabular}}  & 1024 & 48.23 & 51.03 & 51.43 & 0.99 & 0.99 & 0.99 \\
              & 512  & 45.79 & 48.27 & 48.93 & 0.99 & 0.99 & 0.99 \\
\midrule
\multirow{2}{*}{\begin{tabular}[l]{@{}l@{}} \textbf{Fashion-MNIST}\\ \cite{xiao2017fashion} \end{tabular}} & 1024 & 28.60 & 28.76 & 28.66 & 0.99 & 0.99 & 0.99 \\
              & 512  & 28.45 & 28.76 & 28.58 & 0.98 & 0.99 & 0.99 \\
\midrule
\multirow{2}{*}{\begin{tabular}[l]{@{}l@{}} \textbf{CIFAR-10}\\ \cite{krizhevsky2009learning} \end{tabular}}     & 1024 & 35.52 & 37.29 & 36.25 & 0.98 & 0.99 & 0.99 \\
              & 512  & 34.51 & 37.27 & 35.60 & 0.98 & 0.99 & 0.99 \\
\bottomrule
\end{tabular}
}
\vspace{-.75em}
\end{table}

\section{Conclusion}
\label{conc}
%\vspace{-0.25em}
This work presented three CMOS converters that turn the unbiased output of a TRNG into stochastic bit-streams with an arbitrary target probability: block extraction, counter-logic, and a majority-logic pipeline.
All three reach the accuracy of a software random generator, which is the bound for any unpredictable source, and they do so with bits that no algorithm reproduces.
%The proposed methods eliminate SC’s reliance on deterministic LFSR- and Sobol-based generators. They achieve accuracy levels close to software-quality baselines while operating entirely on true-random inputs, providing unpredictability that is not possible with algorithmic sequences.
%providing unpredictability and enhanced security not possible with algorithmic sequences. 
Synthesized in 45nm at 8-bit precision, block extraction and majority logic occupy 72~$\mu$m$^2$ and 61~$\mu$m$^2$ and consume 0.02~pJ and 0.03~pJ per cycle, below LFSR- and Sobol-based generator baselines. 
%Hardware synthesis results demonstrate that the converters offer favorable area–delay–energy trade-offs, particularly for the block-extraction and majority-logic methods, and maintain reliable operation under soft-error injection. 
Under injected soft errors, block extraction degrades no faster than the best deterministic baseline.
The converters are independent of the device technology that supplies the random bits, and each can produce correlated or independent streams by sharing or separating its source.
%The methods are portable across technologies, can be paired with any true-random source, and support both correlated and independent stream generation, offering flexibility for diverse SC applications. 
%proposal can reach software-level error rates, offer favorable area–delay–energy trade-offs, and exhibit stable soft-error behavior. 
%Case studies confirm that the utility of bilinear interpolation slightly surpasses Sobol in PSNR at equal SSIM, and a stochastic-attention mechanism case on the Reuters dataset matches Sobol/LFSR in BLEU score. %IF THERE IS A SPACE -> 
%Future work will focus on co-integrating on-chip TRNGs, extending correlation control mechanism, and deploying the converters in emerging in-memory and edge computing systems.
Together they remove the last deterministic component from an SC datapath and make true-random SC practical on standard CMOS.

%\newpage
%\IEEEtriggeratref{21}
\bibliographystyle{IEEEtran}
\bibliography{Bibliography/References, Bibliography/hassan}

\begin{IEEEbiography}[{\includegraphics[width=1.3in,height=1.25in,clip,keepaspectratio]{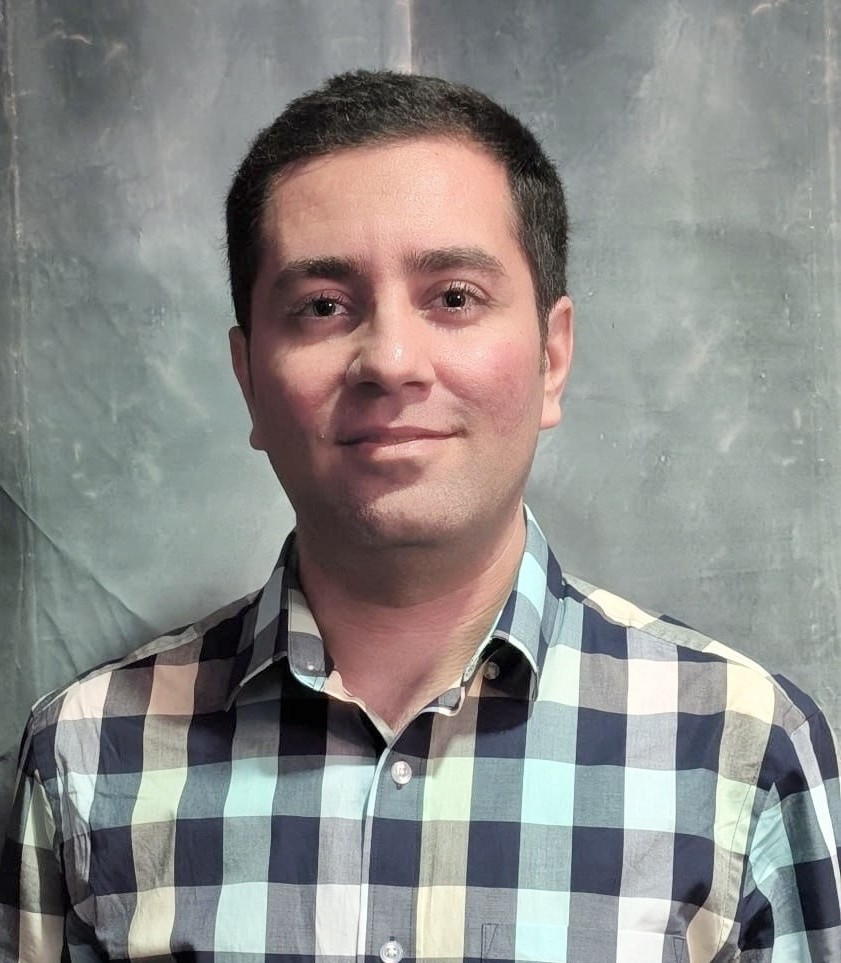}}]{Mehran Moghadam} (S’22) received the B.Sc. degree in Computer Engineering, and M.Sc. degree in \textit{Computer Systems Architecture} from the University of Isfahan, Iran, in 2010 and 2016. He graduated as one of the top-ranking students in both programs.
In 2022, he began his Ph.D. studies in Computer Engineering at the School of Computing and Informatics, University of Louisiana at Lafayette, Lafayette, LA, USA. In 2024, he transferred to the Department of Electrical, Computer, and Systems Engineering at Case Western Reserve University, Cleveland, OH, USA, to continue pursuing his Ph.D. in Computer Engineering.
He became a finalist in the ACM SIGBED Student Research Competition (SRC) at ESWEEK and ICCAD in 2024 and was selected as a Young Fellow in DAC 2024 and 2026. %He has authored/co-authored more than 30 peer-reviewed papers in top EDA venues.
Mehran received the Best Paper Award at ISLPED'26 and the IEEE Circuits and Systems Society (CASS) Pre-Doctoral Grant Award in 2026. 
His research interests include emerging and unconventional computing paradigms, such as energy-efficient stochastic computing models, real-time and highly-accurate hyperdimensional computing systems, bit-stream processing, robust in-memory arithmetic computation, and low-power in- /near-sensor computing designs for edge AI and wearable platforms. 
\end{IEEEbiography}

\begin{IEEEbiography}
[{\includegraphics[width=1.1in,height=1.25in,clip,keepaspectratio]{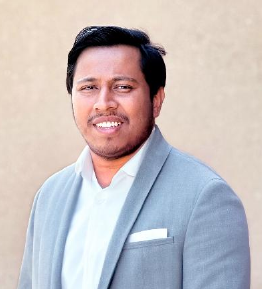}}]{Abu Kaisar Mohammad Masum} (S'25) received the B.Sc. degree in Computer Science and Engineering from Daffodil International University, Dhaka, Bangladesh, in 2020, and the M.S. degree in Computer Science from the University of Louisiana at Lafayette, Lafayette, LA, USA. He is currently a Ph.D. candidate in Computer Science at the University of Louisiana at Lafayette, specializing in quantum machine learning and emerging computing. He was selected as a DAC Young Fellow in 2025 and was a winner of the NASA Beyond the Algorithm Challenge. His research has been published in leading venues, including DAC, DATE, ISLPED, AAAI, TBME, TVLSI, Nature Portfolio, ESL, ISCAS, QCE, ICCD, GLSVLSI, and MWSCAS. His research interests include quantum computing, hyperdimensional computing, hardware-aware AI, and emerging computing systems. He currently serves as the Secretary of the IEEE Computer Society Lafayette Section and is a Student Member of the IEEE.
\end{IEEEbiography}

\begin{IEEEbiography}[{\includegraphics[width=1.3in,height=1.25in,clip,keepaspectratio]{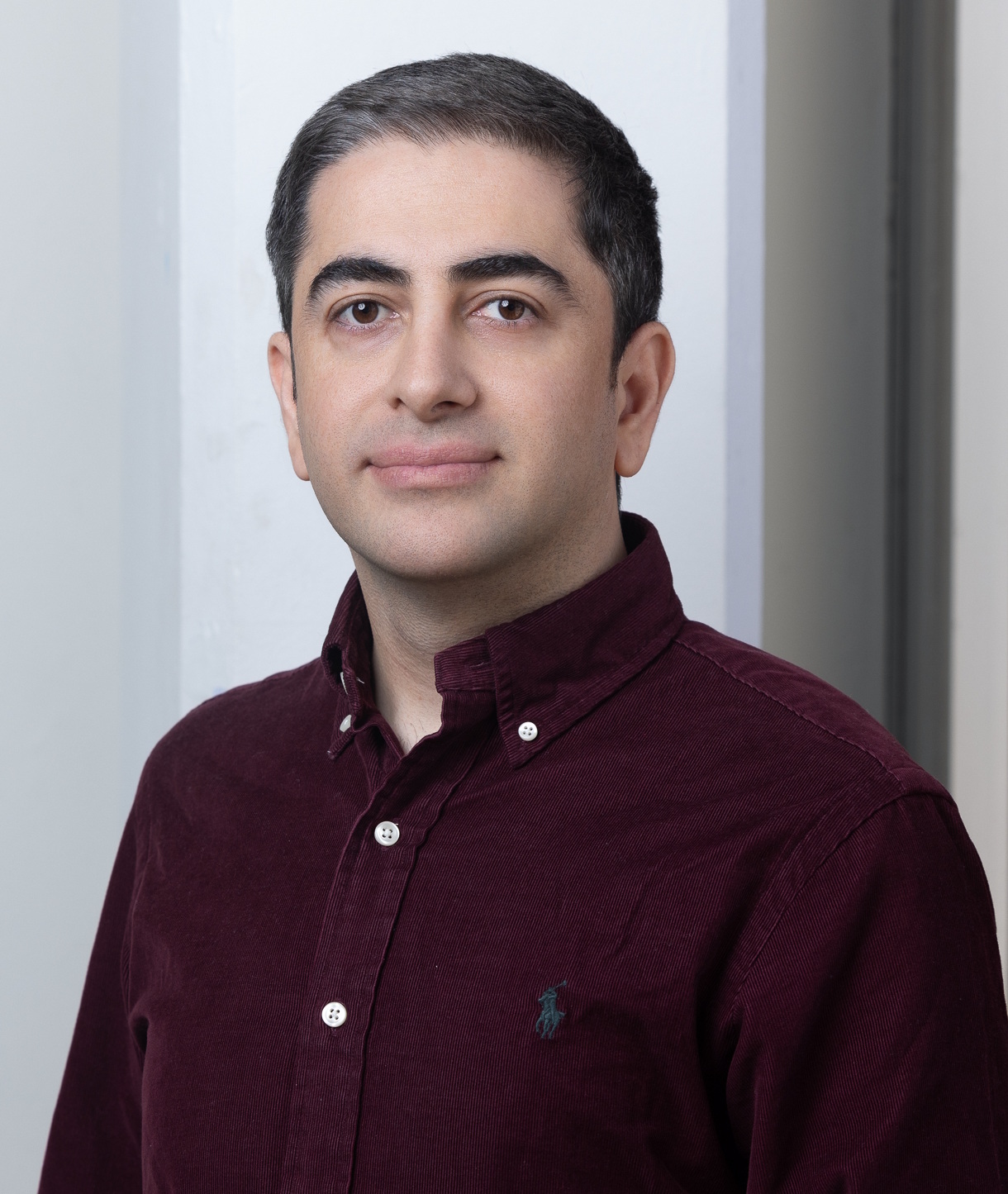}}]{M. Hassan Najafi} (S’15-M’18-SM'23) received the B.Sc. degree in Computer Engineering from the University of Isfahan, Iran, the M.Sc. degree in Computer Architecture from the University of Tehran, Iran, and the Ph.D. degree in Electrical Engineering from University of Minnesota, Twin Cities, USA, in 2011, 2014, and 2018, respectively. He was an Assistant Professor at the School of Computing and Informatics, University of Louisiana at Lafayette, Lafayette, LA, USA, from 2018 to 2024. He is currently an Associate Professor at the Electrical, Computer, and Systems Engineering Department at Case Western Reserve University, Cleveland, OH, USA. His research interests include stochastic and approximate computing, unary processing, in-memory computing, and hyperdimensional computing. He has authored/co-authored more than 120 peer-reviewed papers and has been granted 12 U.S. patents, with more pending. Dr. Najafi received the NSF CAREER Award in 2024, the Best Paper Award at ISLPED'26, GLSVLSI'23, and ICCD’17, the Best Poster Awards at GLSVLSI'24 and DCAS'26, the 2018 EDAA Outstanding Dissertation Award, and the Doctoral Dissertation Fellowship from the University of Minnesota. He has been an editor for the IEEE Journal on Emerging and Selected Topics in Circuits and Systems and a Technical Program Committee Member for many EDA conferences. Dr.~Najafi is a Senior Member of IEEE and a Senior Member of the U.S. National Academy of Inventors (NAI). 
\end{IEEEbiography}

\begin{IEEEbiography}[{\includegraphics[width=1.09in,height=1.2in]{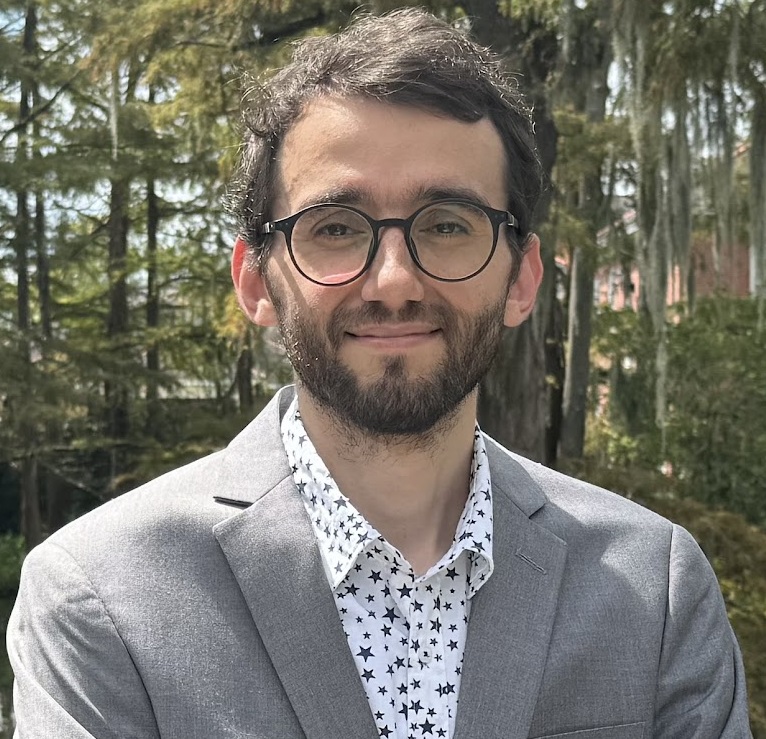}}]{Sercan Aygun} (S’09-M’22-SM'24) received a B.Sc. degree in Electrical \& Electronics Engineering and a double major in Computer Engineering from Eskisehir Osmangazi University, Turkey, in 2013. He completed his M.Sc. degree in Electronics Engineering from Istanbul Technical University in 2015 and a second M.Sc. degree in Computer Engineering from Anadolu University in 2016. Dr. Aygun received his Ph.D. in Electronics Engineering from Istanbul Technical University in 2022. %Dr. Aygun’s Ph.D. work has appeared in several Ph.D. Forums of top-tier conferences, such as DAC, DATE, ASP-DAC, and ESWEEK.
Dr. Aygun received the Best Scientific Research Award of the ACM SIGBED Student Research Competition (SRC) ESWEEK 2022, the Best Paper Awards at ISLPED'26 and GLSVLSI'23, and the Best Poster Awards at GLSVLSI'24 \& DCAS'26. Dr. Aygun's Ph.D. work was recognized with the Best Scientific Application Ph.D. Award by the Turkish Electronic Manufacturers Association and was also ranked first nationwide in the Science and Engineering Ph.D. Thesis Awards by the Turkish Academy of Sciences. He is currently an Assistant Professor at the School of Computing and Informatics, University of Louisiana at Lafayette, Lafayette, LA, USA. He works on tiny machine learning and emerging computing, including stochastic and hyperdimensional computing. 
\end{IEEEbiography}

\begin{IEEEbiography}[{\includegraphics[width=1.09in,height=1.2in]{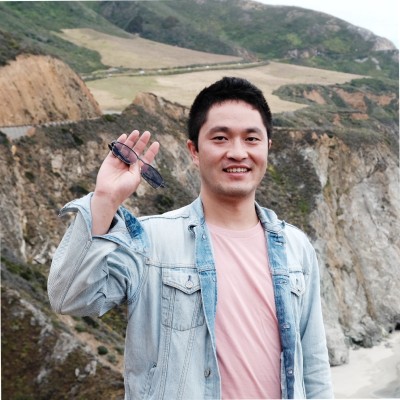}}]{Di Wu} is an assistant professor in the Department of ECE at the University of Central Florida. He earned his PhD from the Department of ECE at the University of Wisconsin-Madison in 2023, with BS and MS degrees from Fudan University. His research interests broadly spread out in emerging areas of computer architecture and systems, such as artificial intelligence, brain-inspired computing, and quantum error correction, featured in ASPLOS, HPCA, ISCA, MICRO, etc. His research has received many recognitions, including one IEEE Micro Top Pick and one Honorable Mention, an inaugural ML \& Systems Rising Star, a dissertation award, a best paper award, an artifact award, and an industry fellowship finalist, etc. His research has received awards from AMD, DOE and NSF.  \end{IEEEbiography}

\vfill

\end{document}